\documentclass[a4paper,fleqn,usenatbib]{mnras}

\usepackage{newtxtext,newtxmath}
\usepackage[T1]{fontenc}
\usepackage{ae,aecompl}

\usepackage{graphicx}	
\usepackage{amsmath}	
\usepackage{amssymb}	
\usepackage{siunitx}
\usepackage[euler]{textgreek}
\usepackage{titlesec}
\usepackage{graphics}
\usepackage{longtable}
\usepackage{tabularx}
\titleformat{\paragraph}
{\normalfont\normalsize\itshape}{\theparagraph}{1em}{}
\titlespacing*{\paragraph}
{0pt}{3.25ex plus 1ex minus .2ex}{1.5ex plus .2ex}
 
\newcommand\numberthis{\addtocounter{equation}{1}\tag{\theequation}}
\newcommand{\SII}{[S\,\textsc{ii}]}
\newcommand{\NII}{[N\,\textsc{ii}]}
\newcommand{\angstrom}{\mbox{\normalfont\AA}}

\title[Extended Red Emission in IC59 and IC63]{Extended Red Emission in IC59 and IC63}

\author[Thomas S. -Y. Lai et al.]{
Thomas S. -Y. Lai,$^{1}$\thanks{E-mail: shaoyu.lai@utoledo.edu}
Adolf N.Witt,$^{1}$
Ken Crawford$^{2}$
\\
$^{1}$Ritter Astrophysical Research Center, University of Toledo, Toledo, OH 43606, USA\\
$^{2}$Rancho Del Sol Observatory, Camino, CA 95709, USA\\
}

\date{Accepted May 2017}

\pubyear{2016}

\begin{document}
\label{firstpage}
\pagerange{\pageref{firstpage}--\pageref{lastpage}}
\maketitle

\begin{abstract}
We analysed new wide-field, wide- and narrow-band optical images of IC 59 and IC 63, two nebulae which are externally illuminated by the early B-star $\gamma$ Cas, with the objective of mapping the extended red emission (ERE), a dust-related photoluminescence process that is still poorly understood, in these two clouds. The spatial distribution of the ERE relative to the direction of the incident radiation and relative to other emission processes, whose carriers and excitation requirements are known, provides important constraints on the excitation of the ERE. In both nebulae, we find the ERE intensity to peak spatially well before the more extended distribution of mid-infrared emission in the unidentified infrared bands, supporting earlier findings that point toward far-ultraviolet (11 eV < E$_\mathrm{{photon}}$ < 13.6 eV) photons as the source of ERE excitation. The band-integrated absolute intensities of the ERE in IC 59 and IC 63 measured relative to the number density of photons available for ERE excitation are lower by about two orders of magnitude compared to ERE intensities observed in the high-latitude diffuse interstellar medium (ISM). This suggests that the lifetime of the ERE carriers is significantly reduced in the more intense radiation field prevailing in IC 59 and IC 63, pointing toward potential carriers that are only marginally stable against photo-processing under interstellar conditions. A model involving isolated molecules or molecular ions, capable of inverse internal conversion and recurrent fluorescence, appears to provide the most likely explanation for our observational results.

\end{abstract}




\section{Introduction}

Extended Red Emission (ERE) is a ubiquitous, photon-driven luminescence phenomenon, present in a wide range of interstellar environments that share one feature in common: the simultaneous presence of interstellar dust and far-ultraviolet (FUV) photons. The intensity of the ERE is closely correlated (albeit not linearly) with the intensity of the FUV radiation field \citep{Smith&Witt02}, ranging over more than four orders of magnitude from the high-latitude diffuse interstellar medium (ISM) (\citealt{Gordon98}; \citealt{Matsuoka11}) to the Orion nebula \citep{Perrin92}. The observational signature of the ERE is a broad (60 nm < FWHM < 120 nm) emission band, variably peaking in the wavelength range 600 nm  <  $\lambda$  <  850 nm \citep{Darbon99} and extending with a red tail of emission well beyond 850 nm. The diffuse nature of the ERE, its relative faintness, its broad spectral extent, and its presence in the company of other, often more intense sources of diffuse continuous emission, such as dust-scattered starlight or recombination continua and line emissions in various types of nebulae, make detecting the ERE a challenging undertaking. For a review of different ERE detection methods, observational results, and derived constraints on likely models for a carrier of the ERE, covering the first two decades of ERE research, we refer the reader to the review by \citet{Witt04}.

Following the discovery of the ERE in the Red Rectangle (\citealt{Cohen75}; \citealt{Schmidt80}), three decades of subsequent research have led to a number of significant observational constraints on the nature of the phenomenon. However, no convincing identification of the carrier of the ERE has emerged, nor has a generally accepted explanation been found for the emission process by which the ERE is produced. 
The only indicator of the likely chemical nature of the ERE carrier comes from the observations by \citet{Furton90} and \citet{Furton92}, showing that ERE is found in carbon-rich but not in oxygen-rich planetary nebulae. This suggests a likely carbonaceous nature of the ERE carrier. Among the most critical constraints on the nature of the ERE carrier are (1) the fact that both the peak wavelength and width of the ERE band are variable but correlated \citep{Darbon99}; (2) the strong correlation between the ERE intensity and the density of the FUV radiation field in the environments in which the ERE is observed \citep{Smith&Witt02};  (3) the apparent absence of ERE in environments whose radiation fields are dominated by stars with T$_{\mathrm{eff}}$ < 10,000K \citep{Darbon99}; (4) a strict lower limit on the photon conversion efficiency of the ERE process of ~(10 $\pm$ 3)\%, assuming that it is driven by the absorption of photons in the wavelength range 540 nm -- 91.2 nm \citep{Gordon98}; and (5) the specific identification of photons with energies E > 10.5 eV as the cause responsible for the ERE in the reflection nebula NGC 7023 \citep{Witt06}. Note that the photon conversion efficiency limit found by \citet{Gordon98} is not an energy conversion efficiency. It is a strict lower limit to the photon conversion efficiency on two accounts: it assumes that all absorbed photons in the 540 nm -- 91.2 nm wavelength range contribute to the ERE excitation and that they are absorbed exclusively by the ERE carriers. Given that other components of the ISM, e.g regular interstellar grains not involved with ERE, as well as molecular hydrogen, do absorb photons in the same wavelength range, and given the likelihood that only far ultraviolet photons are effective in the excitation of the ERE, as suggested by the results of \citet{Darbon99} and \citet{Witt06}, the true photon conversion efficiency of the ERE process could well exceed 100 \% \citep{Witt06}. This imposes a very severe constraint on possible processes responsible for the ERE.

Therefore, more intriguing than the question of a specific carrier for the ERE is the question of the process by which exciting photons are converted into ERE photons, because this latter question may be resolved more readily by accessible observational constraints. There are two fundamentally different processes that are possible mechanisms for the production of ERE: classical fluorescence/photoluminescence \citep{Lakowicz06} and recurrent fluorescence, also known as Poincar\'e fluorescence  \citep{Leger88}. The great majority of ERE models proposed so far rely on classical fluorescence \citep{Stokes1852} or photoluminescence, in which a multi-atom system, e.g. a molecule, molecular ion, nanoparticle, or grain, is excited by absorbing a photon, followed by the possible re-emission of an optical photon of lower energy. The photon conversion efficiency of such processes cannot exceed 100 \%, and is found typically to be much lower than 100 \% in most real systems. Moreover, the energy difference between exciting and emitted photons for optimum photon conversion efficiency, the Stokes shift, is typically of order 1 eV or less. On the other hand, recurrent fluorescence (\citealt{Leger88}; \citealt{Duley09}) in highly isolated molecules or molecular ions relies on inverse internal conversion of vibrational energy into excitation energy of low-lying electronic states (\citealt{Nitzan79}; \citealt{Martin13}; \citealt{Chandrasekaran14}; \citealt{Ebara16}). This process results in the possible emission of more than one photon, following excitation by the absorption of a single photon of sufficiently high energy. The photon conversion efficiency of a given system undergoing recurrent fluorescence increases with the energy of the photon causing the original excitation, and is limited only by the principle of energy conservation and the size of the particle involved. With typical ERE photon energies $\sim$ 1.8 eV and exciting photons from the interstellar radiation field in the diffuse ISM with energy up to 13.6 eV, the Stokes shift for the most efficient recurrent fluorescence lies in the energy range from 9--12 eV. The photon conversion efficiency for this process can readily approach 300 \% for isolated particles, consisting of $\sim$ 20--35 atoms, that are just large enough to resist photo-dissociation in the prevailing interstellar radiation field (\citealt{Leger88}; \citealt{Duley09}). Estimates of a strict lower limit to the ERE photon conversion efficiency of (10 $\pm$ 3)\% \citep{Gordon98} plus several observational indicators suggesting a Stokes shift $\gg$ 1 eV (\citealt{Witt85}; \citealt{Darbon99}; \citealt{Witt06}) seem to favour the Poincar\'e fluorescence mechanism \citep{Duley09}. If this process is responsible for the ERE in ISM environments dominated by atomic hydrogen, the size of the carriers would limited by two conditions. The lower size limit is set by the requirement that particles are just large enough to resist photo-dissociation upon absorption of far-UV photons with energies up to 13.6 eV ($\sim$ 20 carbon atoms, \citealt{Duley09}), while the upper size limit would be set by the balance between cooling via mid-IR vibrational and bending transitions dominant in larger, cooler particles versus cooling by inverse internal conversion, followed by recurrent fluorescence predominant in smaller, therefore hotter, particles. Depending on the assumed heat capacity, the upper size limit may fall in the range from 28 carbon atoms \citep{Duley09} to 36 carbon atoms \citep{Leger88}. It is our objective to investigate this issue in greater detail in the present paper by studying the excitation conditions of the ERE in two externally illuminated reflection nebulae.

\begin{figure}
  \includegraphics[width=\columnwidth]{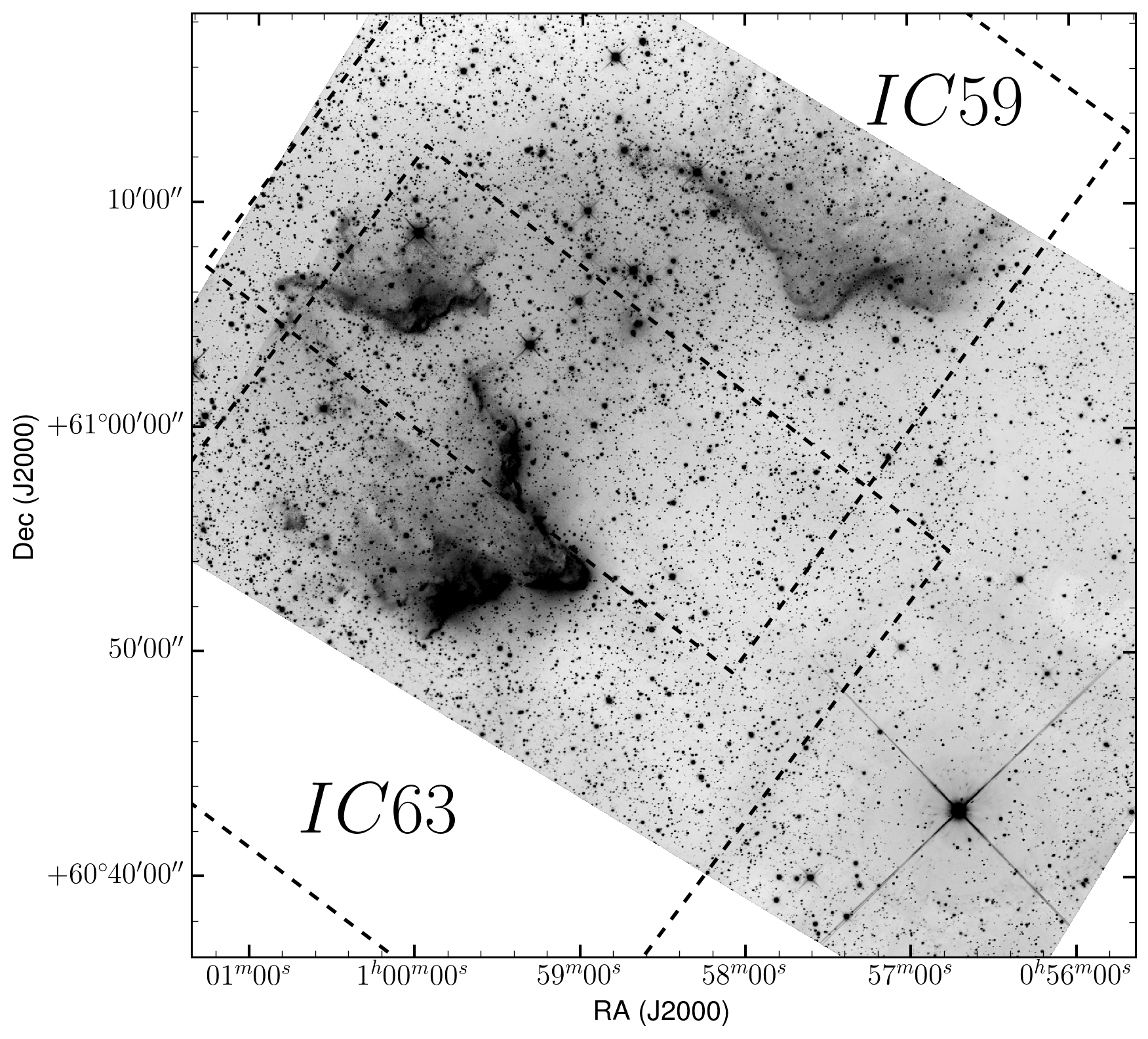}
  \caption{R-band image of $\gamma$ Cas (lower right) and the two nebulae IC 63 and IC 59. The two square dashed outlines delineate the overlapping fields of view covered by the new imaging observations described in Sect. 2. A colour version of this image can be found at \url{http://astrodonimaging.com/wp-content/uploads/2008/08/IC63SROWeb5.jpg}. Image Credit: Don Goldman (reproduced with permission).
}
  \label{Fig:Fig_Geometry}
\end{figure}

The nearby reflection/emission nebulae IC 59 and IC 63 are ideally suited for the study of the excitation of ERE. ERE has been detected previously in the photo-dissociation region (PDR) of IC 63 by spectroscopy \citep{Witt89}, while only upper limits to the ERE intensity are known for IC 59 \citep{Witt90}. Both nebulae are externally illuminated by the early B star $\gamma$ Cassiopeiae  (B0.5 IVpe, d = 168 $\pm$ 4 pc; \citealt{VanLeeuwen07}). The projected positions of the nebulae relative to their illuminating star are shown in Figure \ref{Fig:Fig_Geometry}.

IC 59 has a lower surface brightness than IC 63, which may be the result of both lower column densities as well as a larger physical distance from the illuminating star (\citealt{Blouin97}; \citealt{Karr05}) compared to IC 63. Because of their rare illumination geometry and proximity to Earth, these two nebulae have been studied intensively at wavelengths ranging from the Lyman limit of hydrogen to the hydrogen line and continuum near 21 cm as ideal test cases for PDR models (\citealt{Jansen94}; \citealt{Blouin97}; \citealt{Luhman97}; \citealt{Hurwitz98}; \citealt{Habart04}; \citealt{France05}; \citealt{Karr05}; \citealt{Thi09}; \citealt{Fleming10}; \citealt{Miao10}; \citealt{Perrin92}). As a result of these studies, the physical conditions in these nebulae are exceptionally well constrained.

 The distances of the nebulae from the illuminating star are about one parsec, assuming that they are located near the plane of the sky occupied by $\gamma$ Cas \citep{Friedmann96}. Thus, the incident radiation field at locations within the two nebulae is essentially unidirectional, with the contribution from $\gamma$ Cas about 650 times \citep{Habart04} more intense than the average interstellar radiation field in the solar vicinity at the front PDR in IC 63 facing $\gamma$ Cas. This leads to well-defined PDRs that are viewed essentially edge-on. This is the case particularly in IC 63. Thus, it is possible to observe radiation from different emission processes requiring photons of different energies for their excitation, such as recombination lines from ionized hydrogen and sulphur, fluorescence from photo-excited molecular hydrogen, emissions from ionized and neutral polycyclic aromatic hydrocarbons (PAHs), and dust-scattered continuum at different depths behind the front of the PDR. In this regard, IC 63 is a more ideal geometry for the morphological study of ERE than NGC 7023, where the illuminating star, HD 200775, is centrally embedded and where projection effects make it more difficult to spatially separate different emission processes (\citealt{Witt06}; \citealt{Berne08}). One of the prime objectives of the present study is to determine the spatial distribution of the ERE in IC 59 and IC 63 in relation to other emissions with known excitation requirements.

For this analysis, we take advantage of the wavelength-dependence of the opacity of the interstellar medium consisting of atomic and molecular gas and interstellar dust in the UV-optical spectral range \citep{Ryter96}. An added benefit of the special geometry of IC 59 and IC 63 with respect to $\gamma$ Cas is that scattering by dust occurring here at large scattering angles near 90$^\circ$ is rather inefficient, which makes it easier to separate other UV/optical emission components from the scattered light, including the ERE.

The subsequent chapters of this paper are organized as follows. In Section 2 we present new optical observations of IC 59 and IC 63 and discuss their reductions and absolute intensity calibrations. Additional archival observations obtained from the \textit{WISE}, \textit{Spitzer}, and \textit{Herschel} Space Observatories will be introduced as well. In Section 3 we present the analysis of the ERE morphology in the nebulae based on digital image subtraction and division. We will estimate the ERE intensity using the colour-difference technique and determine the wavelength region of ERE excitation. A discussion of the implications for the ERE emission process and specific ERE carrier models will follow in Section 4, with conclusions to be presented in Section 5. Data tables with the results of measurements of the relative surface brightnesses of IC59 and IC63 are presented in the Appendix.

\section{Observations}
\subsection{Optical Images}
\begin{figure}
  \includegraphics[width=\columnwidth]{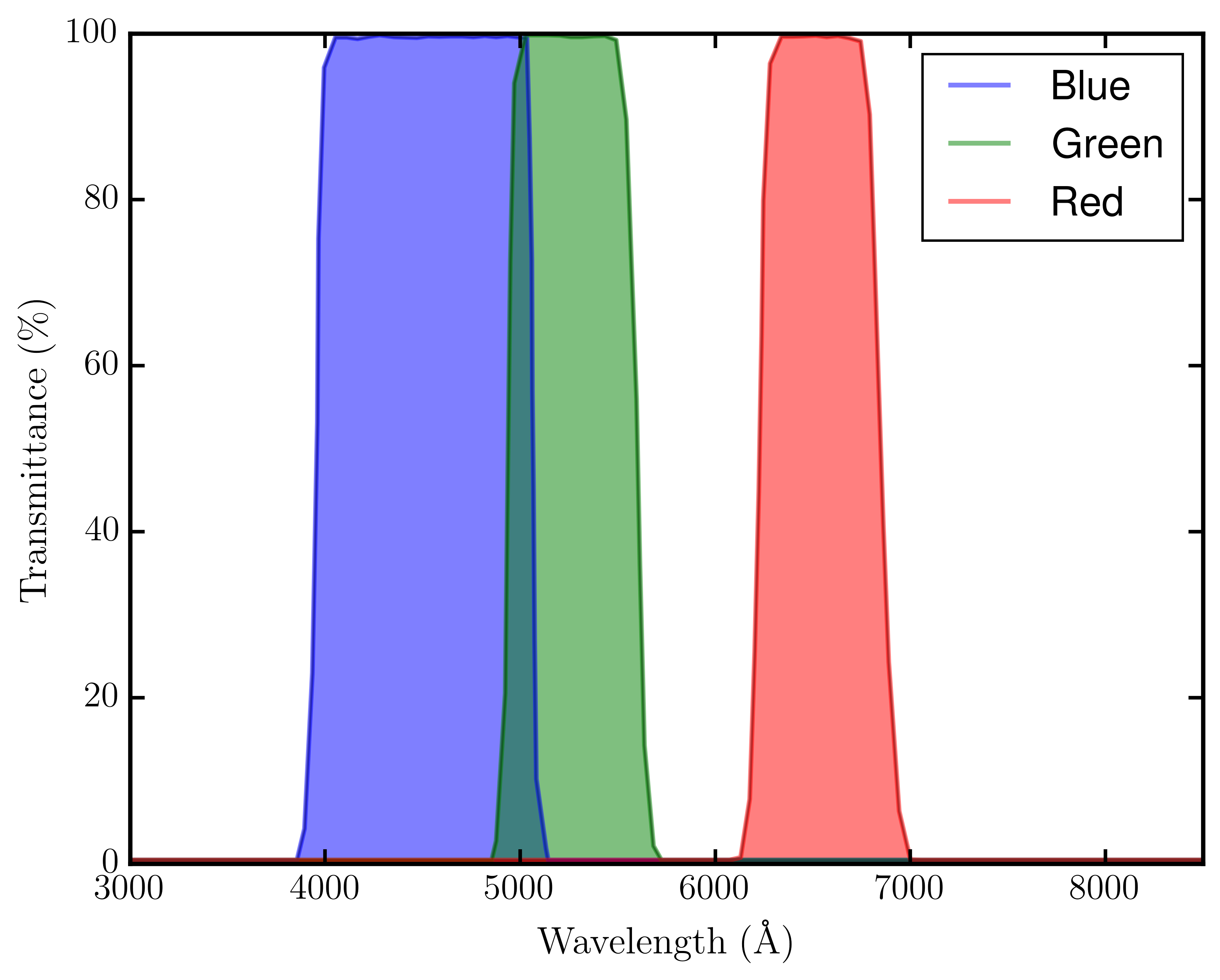}
  \caption{Band passes of the three wide-band Astrodon E-Series filters used in this study. Two additional narrow-band (FWHM = 5.0 nm) filters, centred at 656.3 nm (H$\alpha$ plus \NII \,) and at 672 nm (\SII \,) were used. The latter two filter bands lie within the R-band filter and facilitated the digital image subtraction of nebular line emissions from the scattered light and ERE components.
}
  \label{Fig:Fig_Astrodan_filter}
\end{figure}

\begin{figure}
  \includegraphics[width=\columnwidth]{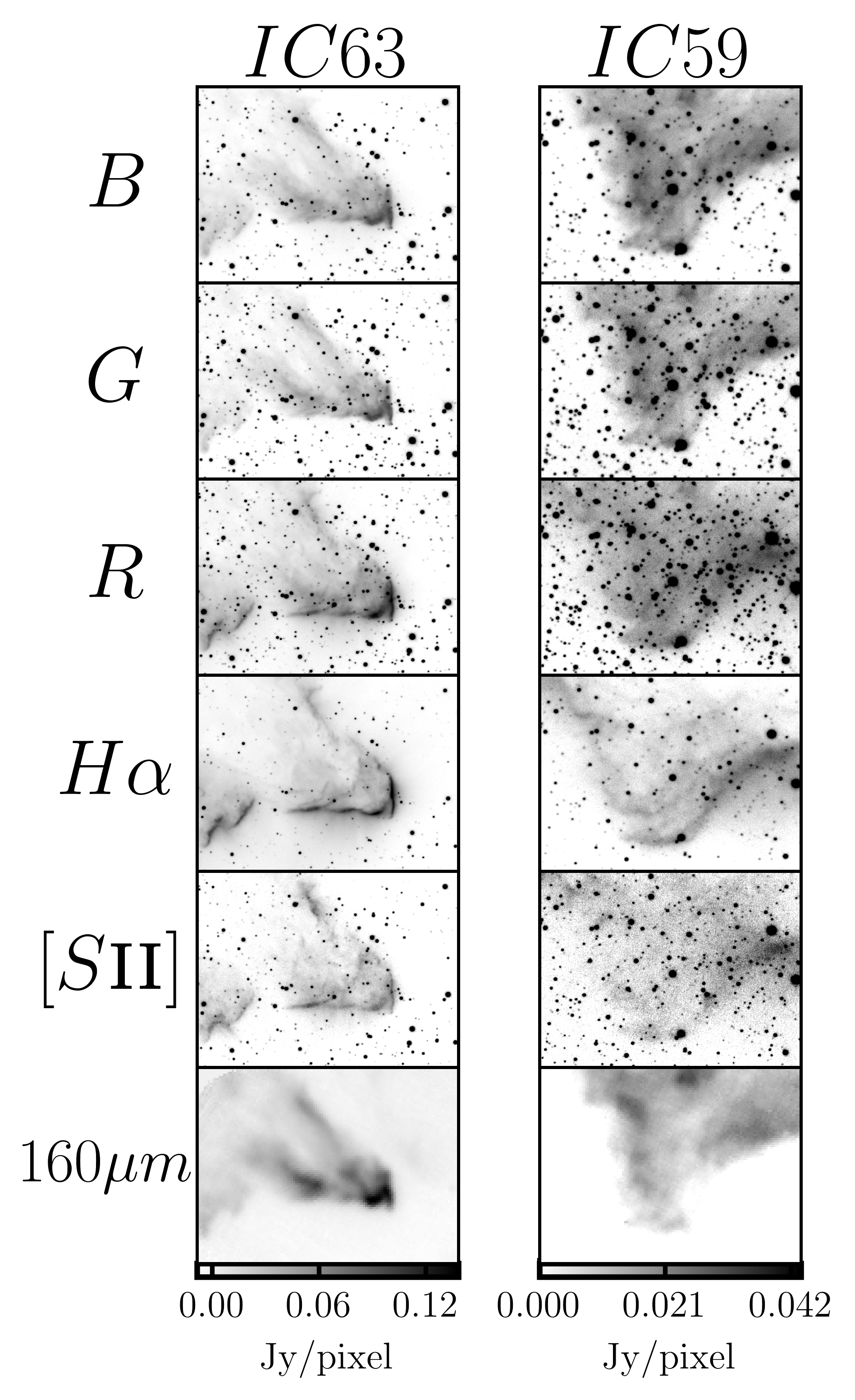}
  \caption{greyscale images of the brightest portion of IC 63 and IC 59 in our five optical bands $B$, $G$, $R$, H$\alpha$, and \SII \, are shown. The bottom panel displays the corresponding image at 160 $\mu$m derived from \textit{Herschel} PACS observations. The intensity scales associated with the bottom panels applies only to these panels. The intensities are measured in units of Jy/pixel, for pixel sizes of 3.2$\arcsec$ $\times$ 3.2$\arcsec$}
  \label{Fig:Fig_Multiband}
\end{figure}

We used the 20-inch RC Optical Systems reflector of the Rancho Del Sol Observatory in Camino, CA, to obtain new, deep images of IC 59 and IC 63. The images were taken during the time period 8-23-2011 to 9-3-2011. The KAF 16803 CCD (4096 $\times$ 4096 px) yielded partially overlapping square (29.3$\arcmin$ $\times$ 29.3$\arcmin$) images, which fully contained the two nebulae and adjacent dark sky but excluded the bright illuminating star $\gamma$ Cas. The image boundaries are indicated in Figure \ref{Fig:Fig_Geometry}. Three Astrodon Generation 2 E-Series filters with nearly square transmission profiles defined our $B$, $G$, and $R$ pass bands, with effective central wavelengths at around 440, 525, and 655 nm. The transmission curves of these filters are shown in Figure \ref{Fig:Fig_Astrodan_filter}. The R-band spectrum of IC 63 of \citet{Witt89} shows the presence of line emissions from H$\alpha$ plus \NII \, and \SII \, \,(unresolved). Therefore, we also imaged both nebulae with narrow-band (5 nm FWHM) filters centred on the H$\alpha$ and \SII \, lines. Information about the exposure times and image centre positions is listed in Table \ref{Tab:Optical_observation}. Image cut-outs centred on the main parts of IC 59 and IC 63 in the five filter bands are shown in Figures \ref{Fig:Fig_Multiband}. A combined, full-colour image of the two nebulae can be found at \url{http://apod.nasa.gov/apod/ap111103.html}.

\begin{table}
	\centering
	\caption{Optical observations}
	\label{Tab:Optical_observation}
	\begin{tabular}{ccc}
		\hline
		Filter & IC63 & IC59\\
		\hline
		$B$ & 10 $\times$ 1200s & 8 $\times$ 1200s\\
		$G$ & 9 $\times$ 1200s & 7 $\times$ 1200s\\
		$R$ & 9 $\times$ 1200s & 8 $\times$ 1200s\\
		H$\alpha$ & 10 $\times$ 1800s & 10 $\times$ 1800s\\
		\SII & 6 $\times$ 1800s & 9 $\times$ 1800s\\
		Image center & $00^{h} 58^{m} 32.5^{s}$ & $00^{h} 59^{m} 29.5^{s}$\\
		& $+61^{d} 12^{m} 53.0^{s}$  & $+60^{d} 53^{m} 59.0^{s}$\\
		\hline
	\end{tabular}
\end{table}

\subsection{Archival Data}
Observations of IC 63 and IC 59 in infrared bands were acquired during the \textit{Spitzer}, \textit{WISE}, and \textit{Herschel} missions. We obtained archival maps in the \textit{Spitzer} IRAC4 band (8 $\mu$m) and in the \textit{WISE} W3 band (12 $\mu$m) for IC 63 and IC 59 from the NASA/IPAC Infrared Science Archive (\url{http://irsa.ipac.caltech.edu/frontpage/}). The \textit{Spitzer} IRAC4 band was designed to be sensitive to mid-IR emissions in bands (7.7 $\mu$m and 8.6 $\mu$m) attributed primarily to ionized PAHs, while the \textit{WISE} W3 covers the wavelength range 7.5 $\mu$m--17 $\mu$m, which includes emissions from both neutral and ionized PAHs (\citealt{Bakes01a}, \citealt{Bakes01b}). Isophotes derived from these maps were used to determine the spatial relationship of the ERE to the emissions attributed to both neutral and ionized PAHs.

In addition to the PAH maps, we used a map of warm and hot molecular hydrogen (H$_{2}$), traced by the emissions in the pure rotational bands H$_{2}$ $\nu$ = 0--0 S(J = 0--5) in the PDR region of IC 63, published by \citeauthor{Fleming10} (\citeyear{Fleming10}, Fig. 7b). These data were obtained with the Infrared Spectrograph (IRS) on the \textit{Spitzer} Space Telescope in spectral mapping mode. Brian Fleming kindly provided us with a fits file of this map. Finally, we acquired far-infrared maps at 160 $\mu$m of both IC 59 and IC 63, obtained with the Photoconductor Array Camera and Spectrometer (PACS) on board of the \textit{Herschel} Space Observatory. The proposer for these observations was Alain Abergel (Proposal \#1342216405 and \#1342216407 for IC 59 and IC 63, respectively). The latter images are displayed in the bottom panels of Figure \ref{Fig:Fig_Multiband}. They were used mainly for the purpose of tracing the spatial distributions of the thermal emission of large dust grains in IC 59 and IC 63 and for assessing the difference between the dust column densities in the two nebulae. The PACS intensity scales are in units of Jy/pixel. The pixel size in the PACS images is 3.2$\arcsec$ $\times$ 3.2$\arcsec$.

\begin{figure*}
  \includegraphics[width=0.85\textwidth]{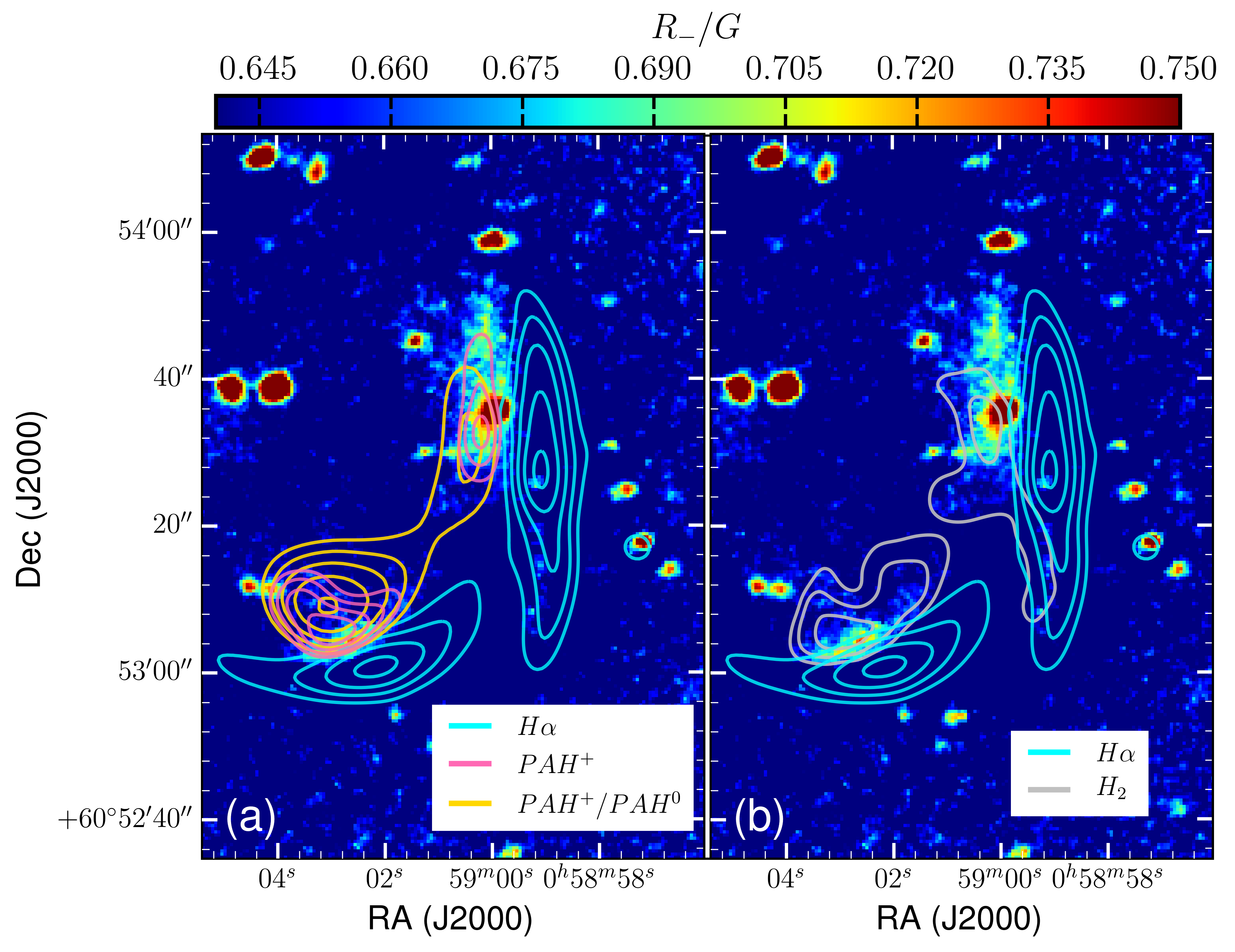}
  \caption{Two of the brightest ERE filaments in IC 63 are revealed by digital division of the R-band image, corrected for H$\alpha$ and \SII \, emission, and the $G$-band image. The exciting radiation arrives from the SW. Both ERE filaments are located immediately behind the hydrogen ionization front traced by H$\alpha$ and in front of emission peaks of ionized PAHs (left panel). The ERE is most closely traced by emission seen in pure rotational lines of H$_{2}$ (right panel).}
  \label{Fig:Fig_Layering_IC63}
\end{figure*}

\section{ERE Morphology and Intensities}

\subsection{Image processing}
In reflection nebulae, the ERE is typically found to be concentrated in bright, narrow filaments, in contrast to the more smoothly distributed scattered light continuum (\citealt{Witt88}; \citealt{Witt89a}; \citealt{Witt89}; \citealt{Witt06}; \citealt{Vijh06}). The spatial distribution of the ERE may, therefore, be seen more clearly by digitally dividing an $R$-band image, containing contributions from both ERE and scattered light, by an image taken in an adjacent continuum band free of ERE, in our case the $G$-band image. However, earlier spectroscopy of IC 63 \citep{Witt89} had revealed the presence of strong H$\alpha$ line emission as well as somewhat weaker \SII \, emission in IC 63. Both these emissions are admitted by our current $R$-band filter and are strongly concentrated in the ionization fronts in IC 63 and IC 59 facing the external stellar source, $\gamma$ Cas (see Figures \ref{Fig:Fig_Multiband}). These contributions must, therefore, be carefully removed from the $R$-band images before the digital division by the $G$-band images can be carried out.

The subtraction of the H$\alpha$ and \SII \, contributions was accomplished in a two-stage process. First, the digital counts in the narrow-band H$\alpha$ and \SII \, images were scaled by the inverse ratio of their exposure times to the R-band exposure times (Table \ref{Tab:Optical_observation}) and corrected for differences in the air masses recorded for all sub-exposures that contributed to the final images. The resulting reduced H$\alpha$ and \SII \, images were then sky-subtracted and subsequently subtracted from the original $R$-band images, assuming an $R$-band extinction coefficient of 0.08 mag/(unit airmass) in accounting for differences in the airmasses between the $R$-band images and the images in the H$\alpha$ and \SII \, filters. In the second stage of the process we applied minor corrections to the subtraction scaling factors to assure that the final sky intensities in regions formerly occupied by the most intense H$\alpha$ ionization fronts blended continuously into sky intensity values of the surrounding sky. These corrections were small, of order two to three percent. The resulting images, designated as $R_{-}$, were then used in the digital division by the $G$-band images, schematically represented by

\begin{equation}
	R_{-}/G = \big\{ R-x[(H\alpha - \mathrm{sky}) + (\SII \, - \mathrm{sky})] \big\}/G
\label{Eqn:R-}
\end{equation}
where $x$ is the scaling factor.

\subsection{ERE Morphology in IC63}
We identified two regions of red excess emission, most likely due to ERE, in the PDR portion of IC 63, facing the exciting star $\gamma$ Cas. These appear as extended bright areas in the $R_{-}/G$ colour maps in Figures \ref{Fig:Fig_Layering_IC63}a,b. In Figure \ref{Fig:Fig_Layering_IC63}a we also display isophotes of the strongest H$\alpha$ emission as well as the isophotes for the peaks in mid-IR emissions seen by IRAC4 and \textit{WISE} W3. The northern ERE region coincides in direction with a $V$ = 16.18 mag background star, identified as star $\# 46$ in \citet{Andersson13}. In order to avoid confusion, we show the same region in Figure \ref{Fig:Fig_Layering_IC63}b with isophotes of emission in pure rotational emission bands of molecular hydrogen observed by \citet{Fleming10}. The H$\alpha$ isophotes are shown again for reference.

Figure \ref{Fig:Fig_Layering_IC63}a reveals that the strongest excitation of ERE occurs immediately behind the ionization front, at the interface with the molecular portion of IC 63. In the northern region, the ERE is nearly coincident with the emission from ionized PAHs seen in the 8$\mu$m band of IRAC but slightly in front of the emission seen in the 12$\mu$m band of \textit{WISE}, which includes bands from both ionized and neutral PAHs. 
In the southern ERE patch, the spatial separation of ionized PAHs and neutral PAHs is more apparent, with the former clearly closer to the ionization front. Here, the ERE is even closer to the ionization front than PAHs in either ionization state.  The spatial distributions of neutral and ionized PAH emission are largely overlapping because the photo-ionization yield of PAHs reaches unity only gradually over an energy range of $\sim$ 9 eV after the ionization potential threshold has been passed (\citealt{Verstraete90}; \citealt{Zhen16}). Consequently, neutral, singly- and doubly-ionized PAHs coexist throughout an extended region behind the hydrogen ionization front, with only gradually changing relative abundances. The fact that the ERE regions are closer to the hydrogen ionization front than PAHs of mixed ionization state suggests that ERE is associated with an environment dominated by PAH cations and, potentially, PAH dications.

Furthermore, Figure \ref{Fig:Fig_Layering_IC63}b suggests a close spatial relationship between warm (T $\sim$ 700 K; \citealt{Fleming10}; \citealt{Thi09}) molecular hydrogen and the ERE. The southern ERE patch shows the ERE to be strongest near the neutral atomic/molecular hydrogen interface, a region characterized by intense pure rotational line emission of H$_{2}$ \citep{Fleming10}. This part of IC 63 has densities n(H$_2$) $\sim (1-5) \times 10^{3}$ cm$^{-3}$ \citep{Thi09} and a column density N(H$_2$) $\sim 6 \times 10^{21}$ cm$^{-2}$ \citep{Thi09}, which is supported by measurements of dust extinction of background stars seen through IC 63 by \citet{Andersson13}. The higher pure rotational levels of H$_{2}$ are most likely pumped by UV photons in the Lyman and Werner bands, which implies an abundance of photons with energies in the range 11.2 eV < E$\mathrm{_{photon}}$ < 13.6 eV in the region where the brightest ERE is observed. A similarly tight spatial correlation between ERE and UV-pumped vibrational emission of H$_2$ was seen in the high-resolution HST observations of NGC 7023 \citep{Witt06}, which further supports this interpretation.

\subsection{ERE Morphology in IC59}
When processed in an analogous fashion, the $R_{-}/G$ ratio image for IC 59 shows no distinct filamentary ERE features. In Figure \ref{Fig:Fig_Ha_W3} we display the $R_{-}/G$ colour map of IC 59 with the superimposed isophotes of H$\alpha$ and \textit{WISE} W3 emissions in the same frame. Note that the isophotes of these two emission processes are largely overlapping spatially when projected onto the sky. This is consistent with a geometry in which the ionization front is not seen edge-on but rather as an extended, inclined face-on surface with H$\alpha$ emission occurring in front of the \textit{WISE} W3 emission, confirming earlier suggestions that IC 59 is located behind the plane of the sky containing the star $\gamma$ Cas (\citealt{Blouin97}; \citealt{Karr05}). The lack of obvious ERE filaments in IC 59 is consistent with this geometry.
This does not imply the absence of ERE in IC 59; it could merely mean that ERE is distributed more nearly uniformly across the face of IC 59. A sensitive method to test for the presence of uniformly distributed ERE is the colour-difference technique \citep{Witt85}, which we will employ in the next section.

\subsection{ERE Detection with Colour Difference Technique}

\subsubsection{Colour Difference Technique}
The colours of a reflection nebula are related to the corresponding colours of its illuminating star through the wavelength dependence of the dust cross sections for scattering and extinction internal to the nebula \citep{Witt85}. If the nebular surface brightness $S$ is a result of scattering alone and if changes in the scattering angle throughout the nebula are small, the colour differences between nebula and star can be approximated as \citep{Witt85}:

\begin{figure}
  \includegraphics[width=\columnwidth]{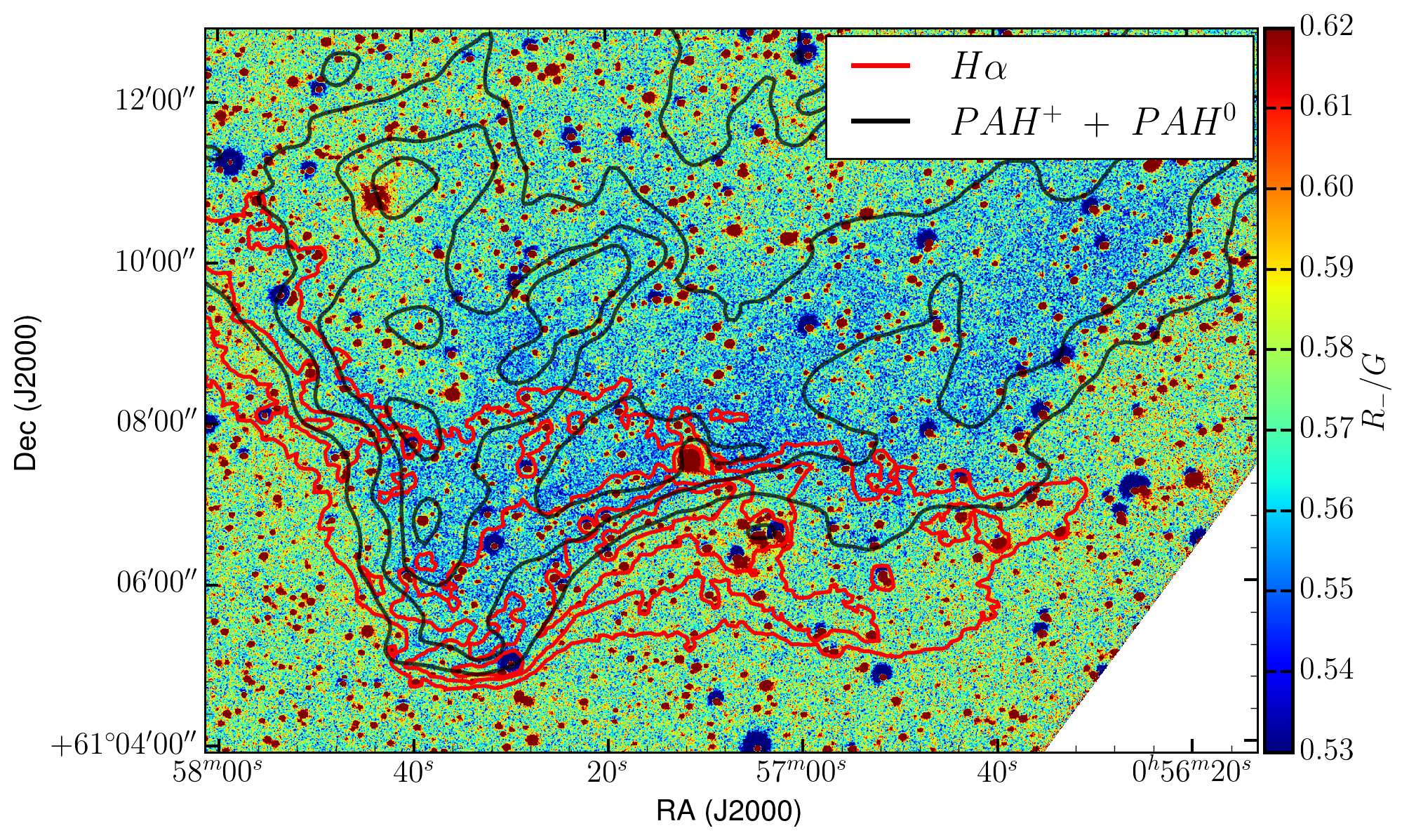}
  \caption{IC 59 $R_{-}/G$ ratio map with \textit{WISE} W3 PAH (black) and H$\alpha$ (red) contours. The H$\alpha$ emission is spatially extended and overlaps the PAH emission in the southern half of IC 59. Combined with the low surface brightness in the scattered light continuum, this morphology suggests that IC 59 is located behind the plane of the sky containing $\gamma$ Cas.}
  \label{Fig:Fig_Ha_W3}
\end{figure}

\begin{align*}
	&\Delta C(B,G) = log(S/F^{*})_B-log(S/F^{*})_G \\
	&\approx log \Bigg\{ \frac{a(B) \big\{ 1-exp[-\tau_0(B)] \big\} exp[\tau_{*}(B)-\tau_1(B)]}{a(G) \big\{ 1-exp[-\tau_0(G)] \big\} exp[\tau_{*}(G)-\tau_1(G)]} \Bigg\} \numberthis
\label{Eqn:ColourDiff_BG}
\end{align*}

\noindent

\begin{align*}
	&\Delta C(G,R_{-}) = log(S/F^{*})_G-log(S/F^{*})_{R_{-}} \\
	&\approx log \Bigg\{ \frac{a(G) \big\{ 1-exp[-\tau_0(G)] \big\} exp[\tau_{*}(G)-\tau_1(G)]}{a(R_{-}) \big\{ 1-exp[-\tau_0(R_{-})] \big\} exp[\tau_{*}(R_{-})-\tau_1(R_{-})]} \Bigg\} \numberthis,
\label{Eqn:ColourDiff_GR}
\end{align*}

where $B$, $G$, and $R_{-}$ are the effective wavelengths of our three wide-band filters, $F^{*}$ is the stellar flux of the illuminating star measured at Earth, $a$ is the dust albedo, $\tau_{0}$ is the total line-of-sight optical depth through the nebula, $\tau^{*}$ is the optical depth of nebular dust in front of the star (if any), and $\tau_{1}$ is the optical depth between the star and the nebular region in which the nebular surface brightness $S$ is being measured. The colour differences $\Delta C(B,G)$ and $\Delta C(G,R_{-})$ are linearly correlated with each other through the wavelength dependences of the dust albedo and optical depths. At small values of $\tau_{1}$ the nebula is typically bluer than the star, with nebular colours becoming gradually redder as $\tau_{1}$ increases, showing the affects of increasing internal reddening of the illuminating starlight.

When ERE is present in the $R_{-}$-band, the colour difference $\Delta C(G,R_{-})$ will be greater than expected for scattering alone, while  $\Delta C(B,G)$ retains its expected value. The deviation of the observed $\Delta C(G,R_{-})$ from the expected value allows us to estimate the fractional contribution made by ERE to the $R_{-}$-band surface brightness.

\subsubsection{Absolute Calibration}

\paragraph{$\gamma$ Cas}
If the images of reflection nebulae include the illuminating stars within the field of view of the camera, the log($S/F^{*}$) data for Eq. (\ref{Eqn:ColourDiff_BG}) \& (\ref{Eqn:ColourDiff_GR}) can usually be obtained by relative photometry \citep{Witt89}, without the need for absolute calibration. In the case of IC 63 and IC 59 we could not follow this approach for several reasons. As shown in Figure \ref{Fig:Fig_Geometry}, the illuminating star, $\gamma$ Cas, is offset from the two nebulae by angular distances of $\sim$20$\arcmin$ and $\sim$25$\arcmin$, respectively, corresponding to a light travel time of 3 to 4 years, and well outside the camera fields centred on the two nebulae. Furthermore, $\gamma$ Cas is a well-known prototype of the class of Be stars, which has exhibited significant brightness and spectral variations on both long and short time scales (\citealt{Henry12}; \citealt{Pollmann14}, and references therein). Our observations of IC 63 and IC 59 were obtained in the late summer of 2011; thus, we need photometric data for $\gamma$ Cas for the period in early 2008. The photometry of \citet{Henry12} yielded values of $V$ = 2.17 and $B-V = -0.19$ for this period. By contrast, SIMBAD values, based on photometry from the 1960$'$s, suggested $V = 2.39$ and $B-V = -0.10$, which would have introduced significant errors.

After deriving transformations between the $B-V$ and $V-R$ colours of the Johnson-Cousins $UBVRI$ system \citep{Bessell90} and the $B-G$ and $G-R_{-}$ colours in our own filter system by convolving the respective band pass functions with a calibrated spectral energy distribution of $\gamma$ Cas, we employed the absolute flux calibration of Vega \citep{Colina96} to arrive at estimates of the absolute flux densities for $\gamma$ Cas in our $B$, $G$, and $R_{-}$ bands that are appropriate for the period of early 2008.  These values are listed in Table \ref{Tab:gammacas}.

\begin{table}
	\centering
	\caption{Flux density for $\gamma$ Cas (early 2008)}
	\label{Tab:gammacas}
	\begin{tabular}{lc}
		\hline
		Band & $F^{*}$\\
		& ($\mathrm{\times 10^{-10} \ ergs \ cm^{-2} \ s^{-1} \ \angstrom^{-1}}$)\\
		\hline
		$B$ & 9.50 \\
		$G$ & 5.83 \\
		$R_{-}$ & 3.40 \\
		\hline
	\end{tabular}
\end{table}

\paragraph{IC63}
We calibrated the images containing IC 63 (Figure \ref{Fig:Fig_Geometry}) with the help of stellar data from \citet{Andersson13}, who provided $V$ magnitudes, MK spectral types, visual extinctions ($A_{V}$), and ratios of total-to-selective extinction values ($R_{V}$) for numerous faint stars seen through and around the nebula. We found the photometry of \citet{Andersson13} stars \#11, \#14, \#15, \#52, \#63, \#68, and \#76 to be an internally most consistent set, and we based our calibration on these stars. We derived apparent magnitudes in our $B$, $G$, and $R_{-}$ bands from the $V$ magnitudes and intrinsic colours corresponding to the MK spectral types, corrected for reddening with values derived from $A_{V}$ and an $R_{V}$-dependent wavelength dependence of extinction from \citet{Cardelli89}. The transformation to absolute fluxes was again based on the absolute flux calibration of Vega \citep{Colina96}. The resulting flux conversion factors are listed in Table \ref{Tab:passbands}.

\begin{table}
	\centering
	\caption{Photometric calibration of our pass bands}
	\label{Tab:passbands}
	\begin{tabular}{lcc} 
		\hline
		Band name & Effective wavelength & Flux per ct/s\\
		& (\angstrom) & $(10^{-17} \mathrm{ergs \ cm^{-2} \ s^{-1} \ \angstrom^{-1}})$ \\
		\hline
		$B$ & 4410 & 2.045\\
		$G$ & 5238 & 2.400\\
		$R_{-}$ & 6464 & 3.530\\
		\hline
	\end{tabular}
\end{table}

\paragraph{IC59}
We transferred the calibration of the IC 63 image to the IC 59 image with the help of suitable nebulosity structure in the overlap region covered jointly by the two images (Figure \ref{Fig:Fig_Geometry}). Small differences were found, likely due to differences in the air masses for individual exposures and night-to-night differences in atmospheric extinction. A sensitive test for a successful cross-calibration was found in the requirement that the measured colour differences for outlying portions of the IC 63 nebula located in the overlap region should yield identical values, regardless of which image was used for the measurements.

\subsubsection{Surface Brightness Measurement}
We measured the sky-corrected surface brightness $S$ in IC 63 and IC 59 in star-free nebular regions in the calibrated $B$, $G$, and $R_{-}$ images, where $R_{-}$ represents the R-band corrected for H$\alpha$ and \SII \,. We normalized the nebular intensities by dividing by the corresponding stellar fluxes of the illuminating star. The results are listed in the Appendix as log($S/F^{*}$) in units of $\mathrm{[sr^{-1}]}$ in Table \ref{Tab:IC59_sbflux} for IC 59 and in Table \ref{Tab:IC63_sbflux} for IC 63, where $F^{*}$ represents the flux values for $\gamma$ Cas for the corresponding bands, listed in Table \ref{Tab:gammacas}. In addition, observed regions are shown in Figures \ref{Fig:Fig_Layering_IC59_W3_IRAC4}a and \ref{Fig:Fig_IC63region} for IC59 and IC63, respectively. Note that the radius of the circular regions are 6.5\arcsec, except for the five ERE brightest patches that are measured by shapes aligned with the filaments (also see Figure \ref{Fig:Fig_IC63_5region}). 

\begin{figure}
  \includegraphics[width=\columnwidth]{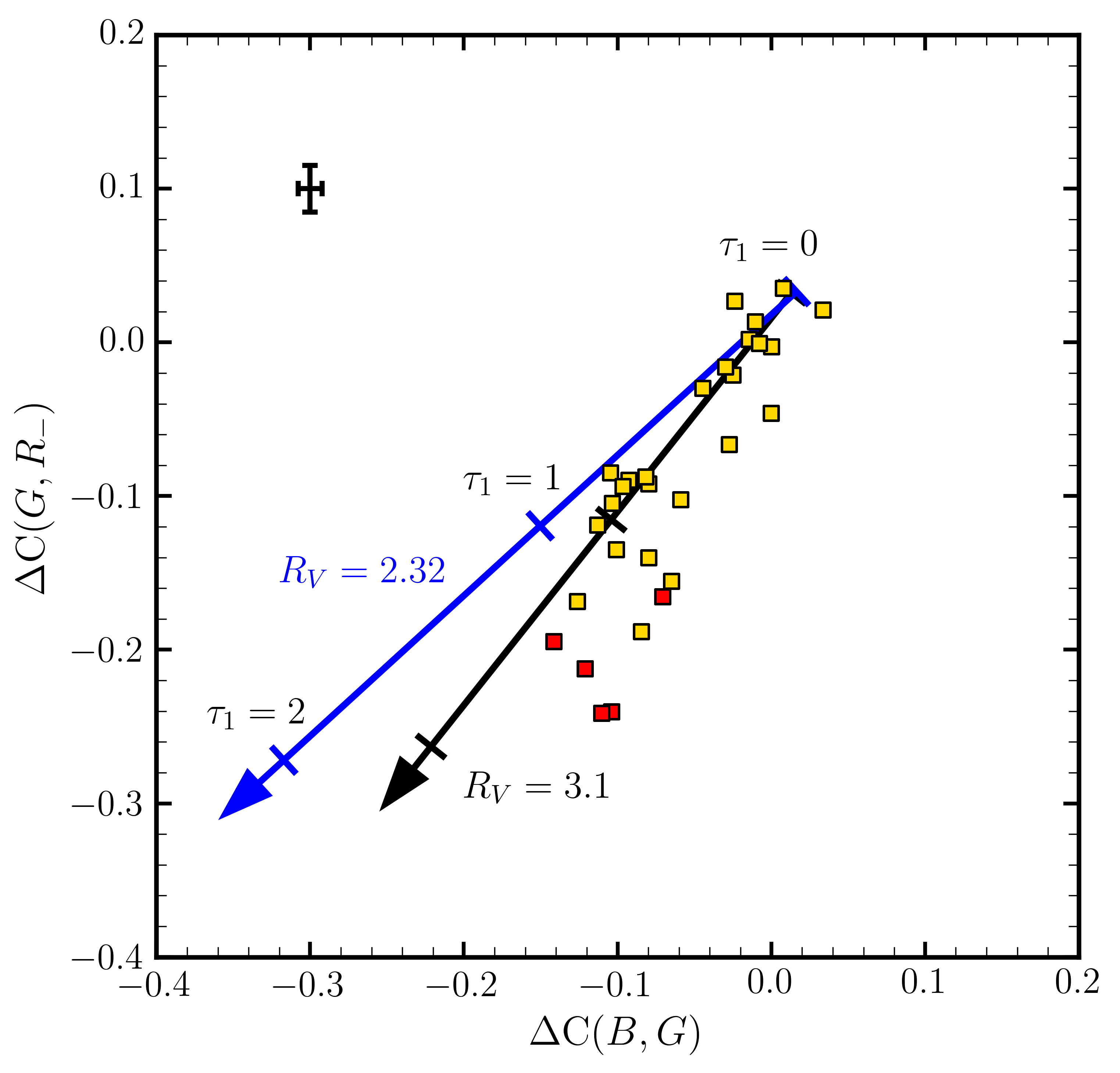}
  \caption{Colour difference diagram for IC 63 (see text for details). Filled squares represent colour differences $\Delta$C($B$,$G$) and $\Delta$C($G$,$R_{-}$) for measured fields in IC 63. The red squares are regions containing the brightest absolute ERE intensities, which correspond to the red squares in Figure \ref{Fig:Fig_Comparison} and the five regions shown in Figure \ref{Fig:Fig_IC63_5region}. The two solid arrows indicate the expected loci for the colour differences for $R_{V}$ = 3.1 (black) and $R_{V}$ = 2.32 (blue), assuming a wavelength-independent dust albedo and no ERE. The typical error bar is about $\pm 0.01$ for $\Delta$C($B$,$G$) and about $\pm 0.02$ for $\Delta$C($G$,$R_{-}$).
}
  \label{Fig:Fig_ColourDiff_IC63}
\end{figure}

\begin{figure}
  \includegraphics[width=\columnwidth]{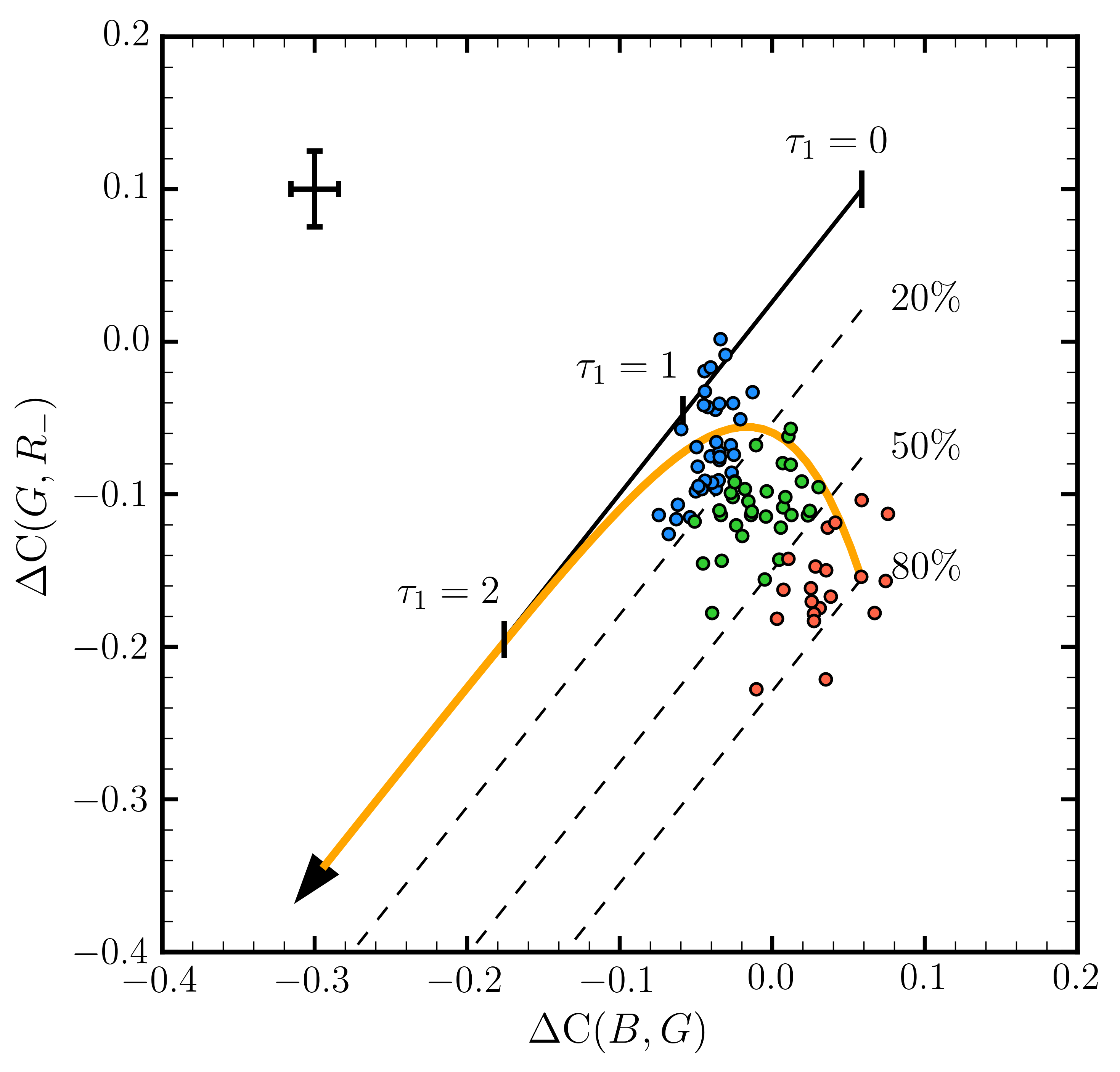}
  \caption{Colour difference diagram for IC 59 (see text for details). Filled circles represent measured colour differences in IC 59, with a map of locations shown in Figure \ref{Fig:Fig_Layering_IC59_W3_IRAC4}. The solid black line is the expected locus for colour differences arising from dust scattered light with a wavelength-independent albedo and no ERE  ($R_{V}$ = 3.1). The yellow solid line represents a model that includes scattered light plus ERE, produced by FUV excitation with E$_{\mathrm{exc}}$ = 10.2 eV. The three dashed lines indicate loci where the ERE intensity  is 20\%, 50\%, and 80\% of the underlying scattered light intensity, respectively. The typical error bar is about $\pm 0.02$ for $\Delta$C($B$,$G$) and about $\pm 0.03$ for $\Delta$C($G$,$R_{-}$).
}
  \label{Fig:Fig_ColourDiff_IC59}
\end{figure}

\subsubsection{Colour Difference Diagrams}
\begin{figure}
  \includegraphics[width=\columnwidth]{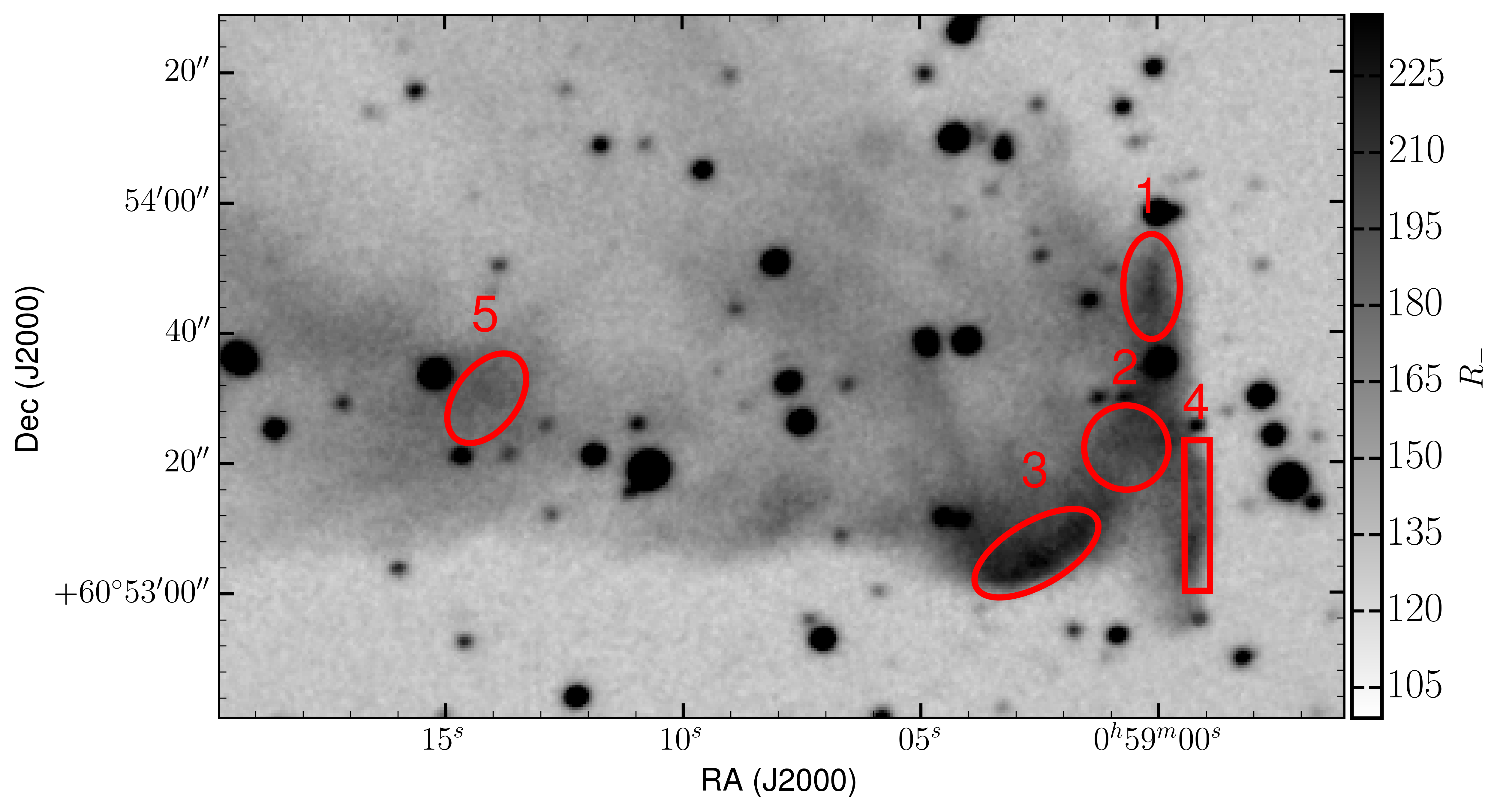}
  \caption{Identification of regions 1,2,3,4,5 in IC 63, the brightest ERE patches originally seen by the image ratio technique (Figure \ref{Fig:Fig_Layering_IC63}), superimposed on a $R_{-}$ map.} 
  \label{Fig:Fig_IC63_5region}
\end{figure}

We computed the colour differences $\Delta$C($B$,$G$) and $\Delta$C($G$,$R_{-}$), as defined in Equations (\ref{Eqn:ColourDiff_BG}) and (\ref{Eqn:ColourDiff_GR}), from the values of the stellar flux-normalized surface brightness values listed in Table \ref{Tab:IC63_sbflux} and \ref{Tab:IC59_sbflux} for IC 63 and IC 59, respectively. The results, respectively, are displayed in Figures \ref{Fig:Fig_ColourDiff_IC63} and \ref{Fig:Fig_ColourDiff_IC59}. In Figure \ref{Fig:Fig_ColourDiff_IC63} we added two model lines for the expected colour differences computed from the first-order scattering approximations in Equations (\ref{Eqn:ColourDiff_BG}) and (\ref{Eqn:ColourDiff_GR}). The black arrow is based on the wavelength dependence  of extinction for standard Milky Way dust with $R_{V}$ = 3.1 of \citet{Weingartner01}, while the blue arrow assumes a value of $R_{V}$ = 2.32  with a wavelength dependence of extinction calculated from \citet{Cardelli89}. The value of $R_{V}$ = 2.32 is an effective average of the extinction characteristics of dust in IC 63 based on the photometry of numerous background stars seen through the nebula, as determined by \citet{Andersson13}. For subsequent steps in the analysis of the data for IC 63 we will adopt the $R_{V}$ = 2.32 model. Both models for IC 63 assume starting values $\tau_0$(B) = 2.0 and $\tau_*$(B) = 0.0, as well as a constant albedo. The direction of the arrows indicates increasing values of $\tau_1$, the optical depth encountered by the illuminating radiation on its path through the nebula. As expected for the presence of ERE, the great majority of the measured regions in IC 63 have $\Delta$C(G,$R_{-}$) colour differences redder than predicted by the $R_{V}$ = 2.32 model. The regions located directly behind the ionization fronts in IC 63, identified in Figure \ref{Fig:Fig_Geometry} as exhibiting peak ERE intensities, are indeed showing the largest deviations from the pure scattering model in Figure \ref{Fig:Fig_ColourDiff_IC63}. These regions are designated as red squares in Figure \ref{Fig:Fig_ColourDiff_IC63}, and are similarly highlighted in Figure \ref{Fig:Fig_IC63_5region} and \ref{Fig:Fig_Comparison}. We discuss the conversion of deviations from pure scattered light models to absolute band-integrated ERE intensities in Section 3.4.5.

The colour difference data for IC 59 (Figure \ref{Fig:Fig_ColourDiff_IC59}) display the classic pattern of ERE excited by FUV radiation \citep{Witt85}. The nebular regions with the bluest (positive) $\Delta$C($B$,$G$)  values exhibit the reddest (negative) $\Delta$C($G$,$R_{-}$) colour differences, indicating that regions illuminated by the least reddened stellar radiation produce the highest percentage of ERE for a fixed intensity of scattered light. The model line for pure scattering, indicated by the black arrow, was computed from Equations (\ref{Eqn:ColourDiff_BG}) and (\ref{Eqn:ColourDiff_GR}), assuming standard Milky Way dust with $R_{V}$ = 3.1. In contrast to conditions in IC 63, we have no indications that the dust characteristics in IC 59 are unusual. The densities inferred from the \textit{Herschel} PACS 160 $\mu$m image (Figure \ref{Fig:Fig_Multiband}) are at least one order of magnitude lower in IC 59 than in IC 63. Consequently, the black model line was computed assuming values for $\tau_0$($B$) = 0.2, reflecting the optically thin structure of IC 59, and $\tau_*$($B$) = 0.0, i.e. no extinction in our line of sight to $\gamma$ Cas. As in the case for IC 63, we assumed a constant albedo in the wavelength range covered by our filters.

In Figure \ref{Fig:Fig_ColourDiff_IC59} we introduced a second model, based on Equation (\ref{Eqn:ColourDiff_GR}) of \citet{Witt85}, which attempts to reproduce the effect of ERE excitation by an external FUV radiation field. While the colour difference $\Delta$C($B$,$G$) was computed as before from Equation (\ref{Eqn:ColourDiff_BG}), Equation (\ref{Eqn:ColourDiff_GR}) for $\Delta$C($G$,$R_{-}$) was modified with the introduction of an ERE term dependent on two parameters: $x$, the maximum ratio of the ERE intensity to the intensity of the underlying scattered light, and $y$, the ratio of the extinction optical depth at the ERE wavelength ($\sim$700 nm) to the absorption optical depth at the wavelength at which the ERE excitation occurs. The model, shown as an orange curve, uses values $x$ = 0.8 and $y$ = 3.6.  The modified equation is

\begin{align*}
	\Delta C(G,R_{-}) \approx & ~ log \Bigg\{ \frac{a(G) \big\{ 1-exp[-\tau_0(G)] \big\}}{a(R_{-}) \big\{ 1-exp[-\tau_0(R_{-})] \big\}}\\ 
	& \times ~ \frac{exp[\tau_{*}(G)-\tau_1(G)]}{exp[\tau_{*}(R_{-})][e^{-\tau_{1}(R_{-})}+x e^{-y\tau_{1}(R_{-})}]} \Bigg\} \numberthis.
\end{align*}

With published optical properties of the Milky Way $R_{V}$ = 3.1 dust model of \citet{Weingartner01}, a value of y = 3.6 points to photons with energies E > 10.2 eV as the source of excitation of the ERE in IC 59, consistent with the results for the nebula NGC 7023 \citep{Witt06}.

\begin{figure*}
  \includegraphics[width=1.0\textwidth]{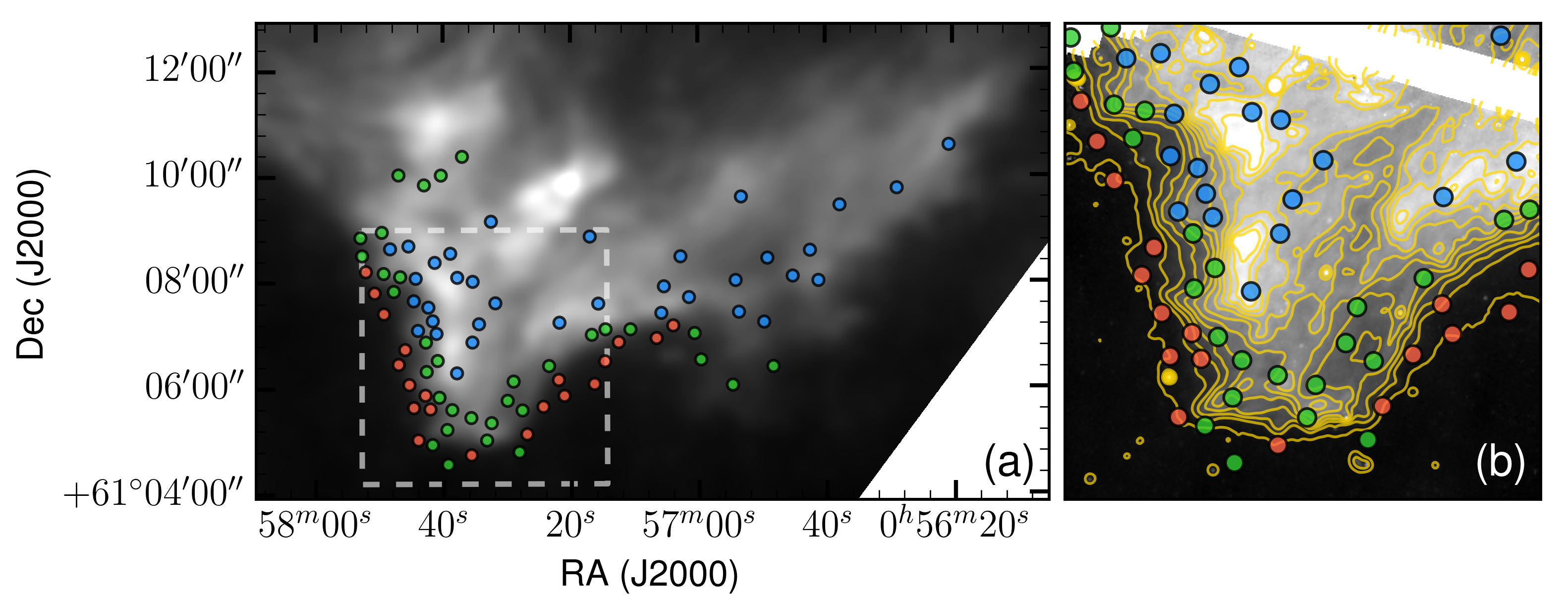}
  \caption{\textit{Left panel}: ERE in IC 59 as a fraction of underlying scattered light, projected on greyscale map of \textit{WISE} 3 intensities. Red circles contain ERE with more than 50\% of the underlying scattered light intensities; green circles contain ERE with 20\% to 50\% of the underlying scattered light intensities; blue circles contain ERE up to 20\% of the underlying scattered light intensities. The greyscale map represents the PAH emission in IC 59 as observed with \textit{WISE} W3. \textit{Right panel}: A region centred on the tip of IC 59, delineated by dashed lines from the lef	t panel, with the greyscale map, overlaid by the isophotes, of the PAH emission observed with \textit{Spitzer} \textit{IRAC4}. The isophotes represent the ionized PAH emission, which extensively overlap with the ERE. The outermost isophote has a value of 3.9 MJy/sr, followed by the next higher isophote with 4.7 MJy/sr.}
  \label{Fig:Fig_Layering_IC59_W3_IRAC4}
\end{figure*}

The three dashed lines in Figure \ref{Fig:Fig_ColourDiff_IC59} represent the loci of nebular regions with ERE intensities of 20\%, 50\%, and 80\% of the intensity of the underlying scattered light continuum, respectively. We colour-coded the regions with ERE > 50\% as red, those with 20\% < ERE < 50\% as green, and those with ERE < 20\% as blue. Figure \ref{Fig:Fig_Layering_IC59_W3_IRAC4}a identifies the positions of the respective regions, superimposed on a greyscale map of IC 59 in the 7.5 $\mu$m -- 17 $\mu$m W3 band of \textit{WISE}, representing the combined emission of neutral and ionized PAHs. Similarly, Figure 
\ref{Fig:Fig_Layering_IC59_W3_IRAC4}b shows a smaller subset of our ERE results superimposed on a \textit{Spitzer} \textit{IRAC4} map of the southern tip of IC 59, which shows the distribution of emission from mostly ionized PAHs. The yellow contours are drawn with a step size of 0.8 MJy/sr. The highest ERE intensities are found in close proximity of the outermost pair of isophotes, corresponding to the range of \textit{IRAC4} intensities from 3.9 MJy/sr to 4.7 MJy/sr. By contrast, the isophotes delineating some of the brightest regions of emission from ionized PAHs in Figure \ref{Fig:Fig_Layering_IC59_W3_IRAC4}b correspond to intensities of about 16 MJy/sr. Clearly, relative to the intensity of the underlying scattered light intensity as well as the emission from neutral and ionized PAHs, the ERE is brightest near the surface facing the illuminating star, demonstrated by the red points. Collectively, bright ERE in IC 59, represented by red and green points, is concentrated in front of the nebular volume containing the PAHs, as seen from the direction of $\gamma$ Cas. 
This is an identical result to that shown for IC 63 in Figure \ref{Fig:Fig_Layering_IC63}. It suggests strongly that the generation of ERE requires higher photon energies than is needed for the excitation of mid-IR PAH features, imaged in the \textit{WISE} W3 band.

\subsubsection{Absolute ERE Intensities}
The vertical displacements from the pure scattering line in the colour difference diagrams yield estimates of the ratio of ERE to scattered light intensity, while measurements of the sky-corrected nebular intensity in the calibrated $R_{-}$ images  of IC 59 and IC 63 provide values of the sum of ERE and scattered light within the $R_{-}$ band. Thus, from the ratio and sum we are able to estimate absolute intensities of ERE and scattered light in the $R_{-}$ band separately. To arrive at band-integrated ERE intensities, we need to take into account that the $R_{-}$ band, after subtraction of the H$\alpha$ and \SII \, bands has only a FWHM of $\sim$50 nm and merely covers the short-wavelength half of the ERE band. The FWHM of the ERE band arising from the illumination of $\gamma$ Cas (log T$_{\text{eff}}$ $\sim$ 4.45) is about 100 nm \citep{Darbon99}, centred near 700 nm, and extending from $\sim$540 nm to 830 nm \citep{Witt89}.

\begin{figure*}
  \includegraphics[width=0.8\textwidth]{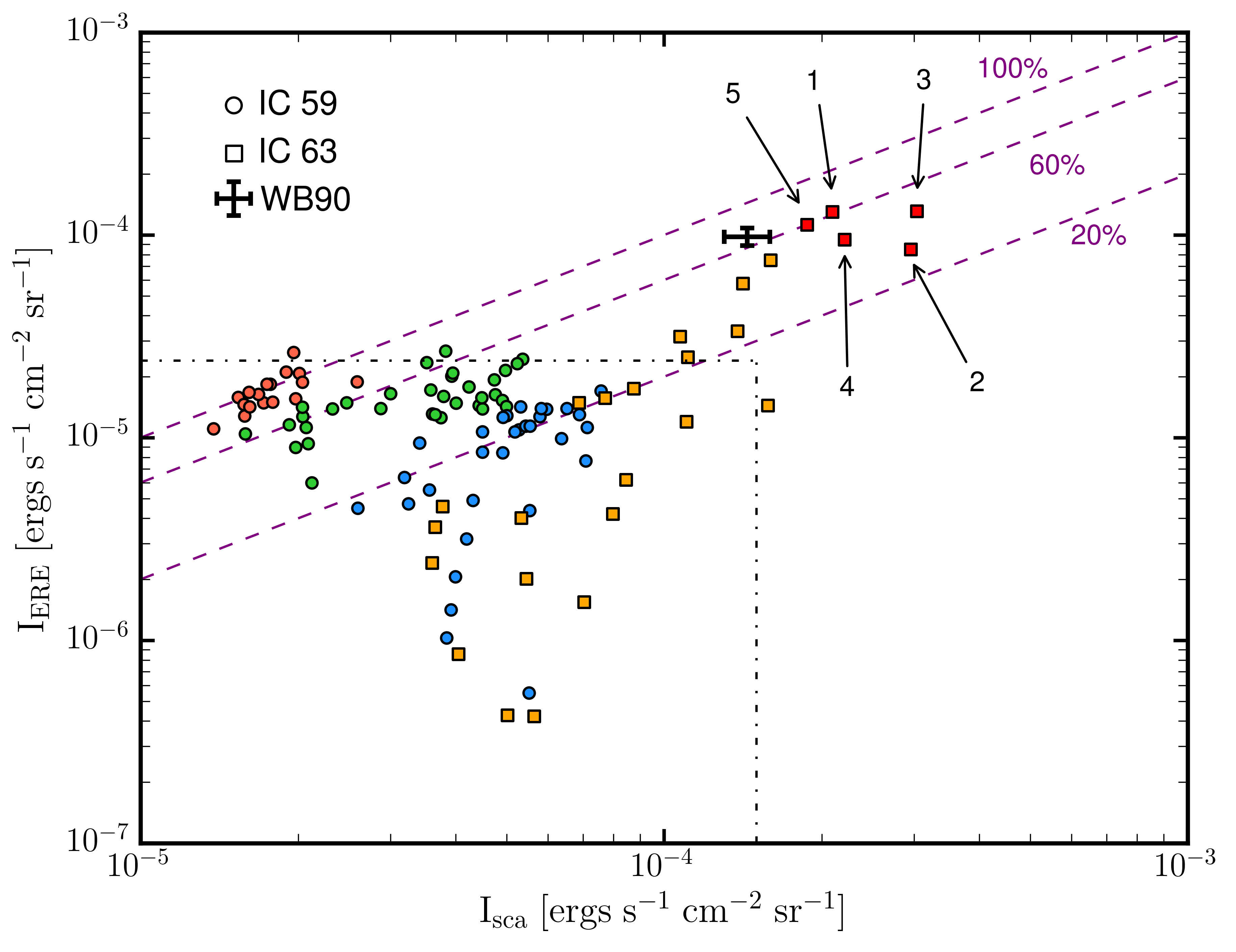}
  \caption{Comparison of absolute ERE intensities as well as absolute scattered light in IC 59 and IC 63. The colour coding for the IC 59 data (filled circles) is identical to that of Figures \ref{Fig:Fig_ColourDiff_IC59} and \ref{Fig:Fig_Layering_IC59_W3_IRAC4}. The horizontal and vertical dashed lines represent the upper limits of the ERE and scattered light intensities in IC 59, previously determined by \citet{Witt90}, using long-slit spectroscopy. The data point WB90 is the only previous detection of ERE in IC 63 \citet{Witt90}. The five brightest ERE regions in IC 63 (red squares) detected in the present work are numbered 1 through 5 and are identified in Figure \ref{Fig:Fig_IC63_5region}. The three diagonal dashed lines indicate locations where the ratio of ERE to dust scattered light intensities is 0.2, 0.6, and 1.0, respectively.}
  \label{Fig:Fig_Comparison}
\end{figure*}

We, therefore, doubled the ERE intensity measured in the $R_{-}$ band to arrive at estimates of total band-integrated ERE intensities. For the corresponding underlying scattered light continuum, we note that the intensity of the scattered light decreases sharply with increasing wavelengths due to the combined effect of the decline of the stellar flux from $\gamma$ Cas with wavelength, coupled with the decrease of the nebular optical depth with wavelength. Thus, we increased the nebular scattered intensity measured in the $R_{-}$ band by a factor of 1.45 to account for the full scattered light intensity underlying the ERE band. These corrections were applied to the measured data from both nebulae. The results are displayed in Figure \ref{Fig:Fig_Comparison} for both nebulae to ease their comparison and discussion.

Two sets of additional dashed lines are added to Figure \ref{Fig:Fig_Comparison} for reference. The black horizontal and vertical dashed-dotted lines represent the upper limits to the band-integrated ERE and scattered light intensities in IC 59, respectively, reported by \citet{Witt90}. Our new observations appear to be fully consistent with these earlier upper limits. Three diagonal dashed red lines represent loci where the percentage of ERE to scattered light has values of 100\%, 60\%, and 20\% respectively. The colour coding of the IC 59 results (coloured circles) is identical to that shown in Figure \ref{Fig:Fig_ColourDiff_IC59} \& \ref{Fig:Fig_Layering_IC59_W3_IRAC4}. We note that the ERE intensities in the red, green, and most of the blue domains fall within a very narrow range from $\sim$ $1 \times 10^{-5}$ -- $2 \times 10^{-5}$ [ergs s$^{-1}$ cm$^{-2}$ sr$^{-1}$], while the scattered light intensities range from $\sim$ $1.4 \times 10^{-5}$ to $7 \times 10^{-5}$ [ergs s$^{-1}$ cm$^{-2}$ sr$^{-1}$]. The large range of values in the ratio of these two intensities is, therefore, largely driven by variations in the scattered light intensity, while the ERE is approximately constant in intensity across the front regions of IC 59. This, now, explains why the digital image division $R_{-}/G$ shown in Figure \ref{Fig:Fig_Ha_W3} did not reveal any ERE filaments.

Maximum ERE intensities and scattered light intensities in IC 63 are larger than corresponding values in IC 59 by roughly identical factors of $\sim$ 6 -- 7. Our new results are in excellent agreement with the only previous ERE detection in IC 63, obtained with spectroscopy with a N-S slit of 7.8$\arcsec$ $\times$ 7$\arcmin$ placed along the PDR, by \citet{Witt89}. Some of the difference between the old and new measurements of the scattered intensity is due to the fact that in the interim $\gamma$ Cas has brightened by at least 0.1 mag in its visual magnitude but probably not in its FUV flux.

Our colour difference technique successfully recovers our results of ERE detection by image division (Figure \ref{Fig:Fig_Layering_IC63}) by showing the ERE patches identified earlier now as regions with the highest ERE intensities. Figure \ref{Fig:Fig_IC63_5region} shows the measurement regions 1 through 5, which are identified with corresponding numbered arrows in Figure \ref{Fig:Fig_Comparison}. The excellent agreement between our results and published earlier detections and upper limits serves as an after-the-fact confirmation that our absolute calibration was reliable (Section 3.4.2.).

The difference in ERE and scattered light intensities between the two nebulae can be explained by the somewhat larger projected distance of IC 59 from $\gamma$ Cas (see Figure \ref{Fig:Fig_Geometry}) and the even larger difference in the physical distances from the star due to a placement of IC 59 behind the plane of the sky containing $\gamma$ Cas and IC 63. We estimate that the physical distance of the front of IC 59 from $\gamma$ Cas is about 2.5 times larger than the distance to the PDR of IC 63 from $\gamma$ Cas. Supporting evidence for this relative spatial arrangement of the two nebulae with respect to $\gamma$ Cas is found in two observational facts. First, as already mentioned in Section 3.3, the H$\alpha$ isophotes of IC 59 are largely overlapping the isophotes of the PAH emission (Figure \ref{Fig:Fig_Ha_W3}), in contrast to IC 63, where the respective isophotes are ordered sequentially with increasing distance from $\gamma$ Cas. Second, as shown in Figure \ref{Fig:Fig_Comparison}, the maximum ratio of ERE to scattered light in IC 59 is about 1.0, while the same ratio in IC 63 is only about 0.6. Assuming that the relative abundances of the ERE carriers in the two nebulae are similar and that the emission of ERE photons is isotropic, this difference is most readily explained with the assumption that the predominant scattering angles for IC 59 are larger than those for IC 63, thus placing IC 59 at a larger distance from earth. This argument relies on the fact that the scattering phase function of interstellar grains is strongly forward directed, leading to a decrease in scattering efficiency with increasing scattering angles.

Finally, a remarkable fact emerges, when we compare the ERE and scattered light intensities in IC 59 with those detected in the high-latitude diffuse ISM by (\citealt{Gordon98}; Fig. 23). The ERE intensities seen in the outer front portions in IC 59 facing the illuminating star (red and green points in Figure \ref{Fig:Fig_Comparison}) are identical to those found in the diffuse ISM with similar dust column densities. If relative abundances of ERE carriers in dusty media are constant, the ERE intensity should scale linearly with the photon density of the exciting radiation field. The density of the FUV radiation field at the PDR in IC 63 has been estimated to be about 650 times that of the interstellar radiation field (ISRF) in the solar vicinity \citep{Jansen96}. With the somewhat larger distance to IC 59 discussed earlier, we estimate that the FUV radiation field incident at the front of IC 59 is approximately 100 times stronger than the ISRF.
We are, thus, led to the conclusion that the relative abundance of the ERE carriers in IC 59 , and by implication, in IC 63, is reduced by about two orders of magnitude compared to their abundance in the diffuse ISM. A possible explanation of this result is that the ERE carriers are marginally stable particles with a distribution of sizes of which only the largest, more stable, particles are able to survive in the intense FUV field produced by $\gamma$ Cas, while the more benign conditions in the diffuse ISM allow for the survival of a much larger number of smaller, less stable entities. This explanation is well supported by the well-established relationship between the peak wavelength of the ERE band and the hardness of the illuminating radiation field (\citealt{Darbon99}; Fig. 1 \& 2). The peak wavelength of the ERE band in IC 63 is near 700 nm, while the ERE in the diffuse ISM peaks near 600 nm \citep{Szomoru98}.

\section{Discussion}
The unique illumination geometries of IC 59 and IC 63 involving the hot B0.5 IVpe star $\gamma$ Cas, coupled with their relative proximity to us (d $\sim$ 168$\pm$4 pc), have enabled us to obtain observations of the spatial distribution and absolute intensity of the ERE in these two objects with unprecedented resolution and sensitivity. We now summarize and discuss the most significant aspects of these observations that affect our understanding of possible models of the ERE excitation process and the likely nature of the ERE carrier.

\subsection{ERE Excitation by FUV Photons}
Digital image division of an $R$-band image containing a partial contribution of ERE in its recorded intensity by an equivalent image in the $G$-band, containing scattered light continuum without ERE, revealed the presence of ERE filaments on the front surfaces of molecular clumps in IC 63 facing the illuminating star. Peak intensities of the ERE (Figure \ref{Fig:Fig_Layering_IC63}) were recorded immediately behind the hydrogen ionization fronts in regions also occupied by UV-pumped rotationally excited molecular hydrogen, but clearly in front of regions dominated by mid-IR aromatic emission features commonly attributed to ionized and neutral PAHs, as observed in the IRAC4 and \textit{WISE} W3 bands, respectively. The emission from rotationally excited molecular hydrogen in IC 63 \citep{Fleming10} is most likely excited by FUV photons in the Lyman and Werner bands, i.e. photons in the energy range 11.2 eV <  E < 13.6 eV. The first photoionization of PAHs begins in the energy range $\sim$ 6 eV <  E < $\sim$ 8 eV (\citealt{Verstraete90}; \citealt{Vijh05}), depending on the size of the PAH molecule, but also requires photons up to E = 13.6 eV in a neutral hydrogen zone to achieve high fractions of ionization \citep{Zhen16}. Given the uni-directional illumination of IC 63, the distance of a particular emission behind the hydrogen ionization front is determined largely by the penetration depth of photons required for the excitation of the respective emission. The location of the ERE peaks relative to the emission regions of molecular hydrogen and ionized PAHs in IC 63, therefore, suggests that ERE is produced with peak efficiency in regions dominated by FUV photons in the 11 eV < E < 13.6 eV energy range. This result is in excellent agreement with the finding of \citet{Witt06} for the nebula NGC 7023, suggesting an energy range 10.5 eV <  E  < 13.6 eV, and is consistent with the conclusions of \citet{Darbon99} that ERE is found only in environments illuminated by stars with T$_{\mathrm{eff}} > 10^{4}$ K, as only those stars can deliver photons in this energy range.

In IC 59, although the detection of the ERE required the use of an alternate technique, i.e. the colour-difference method, the results are fully consistent with those found in IC 63. IC 59 is a much more diffuse nebula compared to IC 63, lacking the optically thick clumps situated closely behind the PDR. Furthermore, IC 59 appears to be located behind the plane of the sky containing $\gamma$ Cas, resulting in an illumination across a broad front rather than in a narrow PDR as seen in a near-edge-on configuration in IC 63. Nevertheless, the rapid change in the ratio of ERE to scattered light with increasing optical depth (Figure \ref{Fig:Fig_ColourDiff_IC59}), the concentration of the ERE in regions well in front of the PAH emission recorded in the \textit{WISE} W3 image (Figure \ref{Fig:Fig_Layering_IC59_W3_IRAC4}a) as well as in front of the intensity peaks of the emission of ionized PAHs recorded by \textit{IRAC4} (Figure \ref{Fig:Fig_Layering_IC59_W3_IRAC4}b), and the almost instantaneous saturation of the absolute ERE intensity in regions of very low optical depth as measured by the scattered light intensity (Figure \ref{Fig:Fig_Comparison}) are all evidence of an excitation process limited to a narrow energy range below the Lyman limit of hydrogen. A formal fit to the IC 59 data in the colour-difference diagram (Figure \ref{Fig:Fig_ColourDiff_IC59}, yellow line) with a simple ERE excitation model \citep{Witt85} resulted in a y-parameter value of 3.6, corresponding to photon energies E > 10.2 eV in the context of a standard $R_{V}$ = 3.1 dust model of \citet{Weingartner01} assumed to apply to IC 59. 

We have independent observational evidence suggesting the presence of dust with $R_{V}$  <  3.1 in IC 63 \citet{Andersson13}, which we took into account in the analysis of the IC 63 data; no such data currently exist for IC 59. Possible differences in the value of $R_{V}$ begin to affect the estimated ERE intensities at larger values of the optical depth $\tau_1$ into the nebula, as illustrated for IC 63 in Figure \ref{Fig:Fig_ColourDiff_IC63}. In the more diffuse nebula IC 59, maximum ERE intensities are already detected near the surface at very small values of $\tau_1$, where possible variations in $R_{V}$ cause only small differences in the inferred intensities of the ERE.

\subsection{Variation of ERE Photon Conversion Efficiency with UV Photon Density}
Compared to most other environments where ERE has been detected \citep{Smith&Witt02}, the system consisting of IC 59, IC 63, and $\gamma$ Cas offers the distinct advantage that the density of the FUV radiation field at the nebular surfaces can be estimated rather accurately. The estimated FUV photon density at the PDR in IC 63 is about 650 times the average galactic FUV photon density \citep{Jansen95}, from which we estimate the FUV photon density at IC 59 to be about 100 times the average Galactic FUV photon density, given the difference in the respective distances from $\gamma$ Cas.

The only other environment with ERE detections, where the radiation field is similarly well known, is the high-latitude diffuse ISM \citep{Gordon98}, where the FUV radiation field density is lower by more than two orders of magnitude than in IC 59. Nevertheless, the absolute ERE intensities in IC 59 and in the diffuse ISM are comparable for similar column densities. We are, thus, led to the conclusion that the relative abundance of the ERE carriers in IC 59, and by implication in IC 63, where maximum ERE intensities are only $\sim$ 4--5 times higher than in IC 59, is reduced by at least two orders of magnitude compared to their abundance in the diffuse ISM. Similarly, a lower-than expected ERE intensity could also indicate a lower photon conversion efficiency per unit mass of ISM in the nebulae compared to the diffuse ISM. A possible explanation of this result is that the ERE carriers are marginally stable particles with a distribution of sizes of which only the largest, more stable, particles are able to survive in the intense FUV field produced by $\gamma$ Cas, while the more benign conditions in the diffuse ISM allow for the survival of a much larger number of smaller, less stable entities \citep{Allain96}. In other words, the average lifetime of the ERE carriers against photo-destruction is likely shorter by at least two orders of magnitude in IC 59 and IC 63, when compared to their lifetime in the diffuse ISM, where the radiation field density is much lower. The shortening of the lifetime of the ERE carriers in the more intense radiation fields is most pronounced for the smallest, least stable particles. Such molecular particles may result by top-down erosion of carbonaceous dust grains in the diffuse ISM (\citealt{Hall95}; \citealt{Duley06}). This explanation is strongly supported by the well-established relationship between the peak wavelength of the ERE band and the hardness of the illuminating radiation field (\citealt{Darbon99}; Fig. 1\&2). The peak wavelength of the ERE band in IC 63 is near 700 nm \citep{Witt89}, while the ERE in the diffuse ISM peaks near 600 nm \citep{Szomoru98}. Assuming that the empirically well-established correlation between the size of molecular particles and the wavelengths of their dominant emission (\citealt{Platt56}; \citealt{Croiset16}) applies to the ERE emitters, the widely observed correlation of the wavelength of peak ERE emission with the density of the prevailing FUV radiation field \citep{Smith&Witt02} would then imply that the smallest ERE carriers responsible for most of the ERE in the diffuse ISM no longer exist in IC 59 and IC 63.

\subsection{Comparison with NGC 7023}
In both IC 63 (Figure \ref{Fig:Fig_Layering_IC63}) and IC 59 (Figure \ref{Fig:Fig_Layering_IC59_W3_IRAC4}), we found the ERE intensity to reach a maximum clearly in front of the emission of both ionized and neutral PAHs, when seen from the direction of the incoming stellar illumination from $\gamma$ Cas. This is contrary to observations in the reflection nebula NGC 7023 by \citet{Berne08} and \citet{Croiset16}, who found extended mid-IR emission attributed to both neutral and ionized PAHs seen in projection well in front of the dominant ERE filament in the NW PDR of NGC 7023, originally mapped by \citet{Witt06}. In contrast to the externally illuminated nebulae IC 63 and IC 59, the illuminating star of NGC 7023, HD 200775, is centrally embedded in the nebula (\citealt{Witt82}; \citealt{Witt92}). While the latter geometry causes some ambiguity about the relative physical distances of the different emission regions from the star, it appears likely that PAH emission does, in fact, occur closer to the star in NGC 7023 than the prominent ERE emission in the NW PDR of that nebula.

We explain the two different sets of behaviour in terms of the different geometries coupled with the different effective temperatures of the illuminating stars. In NGC 7023, the centrally embedded source is a double-lined spectroscopic binary, both components having spectral type B2Ve, corresponding to T$_{\text{eff}}$ $\sim$ 19,000 K \citep{Alecian08}. Optical images of the central region of NGC 7023 suggest the presence of a non-spherical cavity bounded by PDRs, where the stellar flux encounters dense molecular cloud surfaces. These PDRs are locations where strong molecular hydrogen emission and bright ERE are observed \citep{Witt06}. In the cavity interior, molecular hydrogen is photo-dissociated and converted into atomic hydrogen. HD 200775 is not hot enough to ionize significant amounts of hydrogen, allowing atomic hydrogen to serve as a shield that prevents photo-destruction of PAHs and C60 by stellar Lyman continuum photons \citep{Croiset16}, which are not present in significant numbers in the first place. This permits these species to exist in the cavity's interior and to respond to the high density of exciting photons. ERE carriers, assumed to be smaller and thus less stable against photo-destruction, survive mainly near the PDRs, where they emerge from the molecular cloud, and they are then progressively destroyed at a higher rate as they stream away from the PDR into the nebular cavity. The relative spacing of PAHs relative to ERE is interpreted as a result of different degrees of stability against photo-destruction in an atomic hydrogen environment, pervaded by a modestly hard ultraviolet radiation field.

In IC 63, where projection effects are resolved with the least ambiguity, we observe a very hard UV radiation field from an external T$_{\text{eff}}$ $\sim$ 29,000 K star \citep{Telting93}, $\gamma$ Cas, impinging on a dense molecular cloud front. The gas in front of the PDR, facing the star, is fully ionized. No ERE nor PAH emission are observed in this zone, suggesting that neither PAHs nor ERE carriers survive long in the ionized region, with hot electrons and Lyman continuum photons prevailing there. In the zone immediately behind the hydrogen ionization front, hydrogen exists mainly in atomic form as a result of photodissociation of hydrogen molecules by photons in the energy range 11.2 eV -- 13.6 eV. These photons are also ideal for the excitation of the ERE, and they are readily available, as long as hydrogen is predominantly atomic. This accounts for our observation of peak ERE intensities immediately behind the hydrogen ionization fronts in IC 63 (Figure \ref{Fig:Fig_Layering_IC63}). As hydrogen becomes increasingly molecular deeper in the interior of IC 63, the rising molecular hydrogen opacity becomes a serious competitor for photons able to excite the ERE, thus quenching it. Photons capable of ionizing ($\sim$ 6 eV -- 11 eV) and exciting ($\sim$ 4 eV -- 11 eV) PAHs are able to stream deeper into the molecular interior of IC 63, limited only by dust opacity. Thus, the relative spacing in distance of different emission processes in the clumpy cloud interior behind the ionization front is controlled by the nebular opacity for the photons needed to excite the different emissions. As explained earlier, the spatial distributions of the emission from ionized and neutral PAHs will always be largely overlapping, because the ionization yield of PAHs increases only gradually with energy, once the ionization potential has been exceeded (\citealt{Verstraete90}; \citealt{Zhen16}). Therefore, ionized and neutral PAHs largely coexist spatially, with the degree of ionization gradually declining with depth, well behind the region with maximum ERE intensity . This is confirmed by the observational data shown in Figure \ref{Fig:Fig_Layering_IC63}.

Another case where ERE is observed well in front of PAH band emissions is the compact H II region Sh 2-152 \citep{Darbon00}. The embedded location of the ionizing O 9V star leads to projection ambiguities, resulting in a distribution in which ERE and H$\alpha$ emission appear to be roughly co-spatial in the inner part of the nebula, followed by emissions in various PAH bands at greater distances in the outer parts of the nebula. We interpret the apparent superposition of ERE and H$\alpha$ emission as the projection of an inner volume filled with ionized hydrogen, surrounded by a thin, continuous shell source of ERE. As in IC 63, the more distant molecular gas still receives lower-energy photons capable of ionizing and exciting PAHs.

In IC 59, which appears to be located about 2 pc behind the plane of the sky containing $\gamma$ Cas and IC 63, we are observing distinct emission layers largely overlapping by projection. The H$\alpha$ isophotes (Figure \ref{Fig:Fig_Ha_W3}) substantially overlap the PAH emission, appearing only slightly closer to the star. Similarly, the ERE extends over large parts of IC 59 (Figure \ref{Fig:Fig_Layering_IC59_W3_IRAC4}), but reaches its peak intensity level well in front of the PAH emission.

\subsection{The ERE Emission Process and ERE Carrier Models}
The ultimate goal of ERE studies has always been the identification of the ERE carrier. A variety of ERE models has been proposed in the past to account for the existence of ERE, but none of them has been met with lasting acceptance. In Table \ref{Tab:ere_models}, we list the principal ERE models to ease their comparison. There are three fundamental mechanisms that can explain the production of the ERE while meeting at least two of the observational constraints: (i) Classical fluorescence/photoluminescence, (ii) Two-step processes, and (iii) Recurrent fluorescence (RF). All three categories of processes rely on photons as the source of excitation and all aim to produce an emission band similar in width and wavelength as that of the ERE band (\citealt{Witt90}; \citealt{Darbon99}).

\begin{table*}
	\centering
	\caption{ERE models}
	\begin{tabular}{ll} 
		\rule{0pt}{4ex}
		\textbf{I. Classical Fluorescence/Photoluminescence} & References \\
		\hline
		\hline
		Hydrogenated Amorphous Carbon & \citet{Duley85}; \citet{Godard10}\\
		Polycyclic Aromatic Hydrocarbons & \citet{D'Hendecourt86}\\
		Quenched Carbonaceous Composite	& \citet{Sakata92}\\
		C$_{60}$ & \citet{Webster93}\\
		Carbon Clusters & \citet{Seahra99}\\
		Silicon Nanoparticles & \citet{Ledoux98}; \citet{Witt98}\\
		Biofluorescence & Hoyle \& Wickramasinghe (\citeyear{Hoyle96},\citeyear{Hoyle99})\\
		MgSiO$_3$ Silicate & \citet{Thompson13}\\
		Nanodiamonds & \citet{Chang06}\\
		\hline
		\rule{0pt}{4ex}
		\textbf{II. Two-Step Processes}\\
		\hline
		\hline
		PAH Di-cations & \citet{Witt06}\\
		PAH Dimer Cations & \citet{Rhee07}\\
		\hline
		\rule{0pt}{4ex} 
		\textbf{III. Recurrent Fluorescence in Isolated Molecules}\\
		\hline
		\hline
		Poincar\'e Fluorescence & \citet{Leger88}\\
		Thermally Excited Molecules & \citet{Duley09}\\
		\hline
	\end{tabular}
	\label{Tab:ere_models}
\end{table*}

The first category of models, relying on classical fluorescence/photoluminescence, identifies materials that are considered more or less likely to exist in interstellar space and that have been shown experimentally to exhibit fluorescence/photoluminescence in the red spectral range with a band similar in width as the ERE. A common shortcoming of these models is that they exhibit maximum fluorescence efficiencies, when the Stokes shift, the energy difference between exciting and generated photons, is relatively small, generally less than 1 eV. This is clearly contradicted by our results for IC 63 and IC 59, shown in Section 3, which consistently identify FUV photons with energies of $\sim$ 10 eV and above as the source of ERE excitation in these nebulae, further confirming similar results by \citet{Witt06} and \citet{Darbon99} for other nebulae. Furthermore, the maximum theoretically possible photon conversion efficiency of classical fluorescence/photoluminescence processes is 100\%. If the energy range of the exciting photons is limited to the FUV, i.e. $\sim$ 11 eV to 13.6 eV in non-ionized nebulae and the diffuse ISM, this limit on the photon conversion efficiency makes it nearly impossible to account for the ERE intensities observed in the diffuse high-latitude ISM \citep{Gordon98} and in the Red Rectangle \citep{Witt06}. It should be noted that even under optimal laboratory conditions the materials proposed as potential ERE carriers (Table \ref{Tab:ere_models}, Category 1) do not even approach fluorescence photon conversion efficiencies of 100\%.

The two-step processes proposed by \citet{Witt06} and \citet{Rhee07} were designed specifically to overcome the shortcomings of the classical fluorescence/photoluminescence models, while relying upon materials, PAHs and PAH dimers, believed to be present in the ISM with substantial abundances. Both models in this category see the role of FUV photons as the necessary tool for the preparation and maintenance of a population of ERE carriers, by the double ionization of PAHs to the dication stage or the ionization of PAH dimers, respectively. Thus, the morphology of the ERE is determined by the presence of the FUV photons required for these processes. This aspect of the models is supported well by the observed ERE morphology in IC 63 (Figure \ref{Fig:Fig_Layering_IC63}) and in IC 59 (Figure \ref{Fig:Fig_Layering_IC59_W3_IRAC4}). The subsequent ERE production still depends on classical fluorescence, which can utilize the full optical/near-UV incident spectrum for excitation, thanks to the abundance of electronic transitions thought to be available in these spectral ranges in either PAH dications or singly-ionized PAH dimers. The biggest shortcoming of these two models is the lack of experimental data on representative optical spectra and fluorescence photon conversion efficiencies, the expectations having been based solely on quantum-chemical calculations in both cases. PAH dications may face the additional difficulty of possible instability due to Coulomb explosion, although this may not be as severe a limitation as frequently thought \citep{Rosi04}.

RF \citep{Leger88} was one of the first models proposed to explain the ERE. Initially introduced as Poincar\'e fluorescence, this mechanism relies upon the intra-molecular process of inverse internal conversion, which follows the initial absorption of a FUV photon and subsequent rapid internal conversion in highly isolated polyatomic molecules or molecular ions \citep{Nitzan79}. Inverse internal conversion turns an initially large reservoir of vibrational energy into electronic excitation in systems where low-lying electronic states exist. Only recently has it been possible to experimentally confirm RF by measuring the critical rate coefficients in C$_{6}^{-}$ (\citealt{Chandrasekaran14}; \citealt{Martin13}) and by observing the resulting red photons \citep{Ebara16}. It is an intrinsic characteristic of RF that its expected photon conversion efficiencies increases with a growing energy difference between that of the exciting photon and the resultant fluorescence photons, a behaviour totally opposite to that seen in classical fluorescence or photoluminescence. With sufficiently high excitation energies, provided in the ISM by FUV photons, RF, as its name suggests, can provide multiple fluorescence photons from a single excitation, resulting in potential photon conversion efficiencies of several hundred percent (\citealt{Leger88}; \citealt{Duley09}). The highest photon conversion efficiencies with a given FUV excitation is produced by the least-massive molecules or molecular ions that are still marginally stable in the given radiation environment. Thus, the RF model for the ERE provides a simple and very direct explanation of two critical facets of the ERE phenomenon, the need for FUV photons to excite (\citealt{Darbon99}; \citealt{Witt06}) and the exceptionally high photon conversion efficiency \citep{Gordon98}. We note that the RF model predicted these two aspects of the ERE well before they were confirmed observationally. Finally, the apparent decrease in the abundance of ERE carriers in IC59 and IC63 relative to the diffuse ISM is also consistent with a model in which the highest photon conversion efficiency is associated with the smallest, least stable carrier particles, as predicted by RF.

\citet{Ito14} demonstrated that only some polyatomic molecular systems engage in RF while other do not. The presence of low-lying electronic states appears to be a necessary requirement for RF. The viability of the RF model for the ERE does, therefore, depend on the presence in the ISM of polyatomic molecular systems with low-lying electronic states. Observations of diffuse interstellar bands (DIBs) (\citealt{Hobbs08}, \citealt{Hobbs09}) suggest that there are well over 500 species of (mostly) unknown molecular systems present in the ISM, with transitions strongly concentrated in the 540 nm -- 760 nm wavelength range, the same wavelength range occupied by the ERE. These systems would satisfy the requirement for the occurrence of RF. It is, therefore, possible that the same population or partially overlapping populations of molecular particles are responsible for DIBs via electronic absorption from the ground state as well as for the ERE via RF, following excitations by FUV photons \citep{Witt14}. 

The apparent high abundance of molecular systems with low-lying electronic states in the ISM should not come as a surprise. By enabling RF on a rapid timescale \citep{Chandrasekaran14}, these systems are able to undergo highly efficient cooling after absorbing FUV photons, thus avoiding other energy-dissipating pathways such as fragmentation or ionization. Those pathways are more likely to occur in molecular systems not capable of RF. Such systems can use only relatively slow mid-infrared transitions for radiative cooling, exposing them for a more extended period during which photo-fragmentation might occur. Thus, RF provides a positive survival benefit for otherwise marginally stable molecular systems in the ISM. 

Our analysis of the new ERE observations in IC 63 and IC 59 reported here has provided strong new evidence that ERE excitation is the result of absorption of FUV photons. Furthermore, the comparison of the band-integrated ERE intensities in these two nebulae with those seen in the diffuse ISM at high galactic latitudes supports the suggestion that ERE carriers are only marginally stable in the diffuse ISM and are subject to deactivation, most likely by photo-fragmentation, in regions of strongly enhanced FUV fields such as encountered in IC 63 and IC 59. These two aspects of the ERE are difficult to match by any of the classical luminescence models listed under category 1 in Table \ref{Tab:ere_models}. On the other hand, the RF model does provide a natural explanation for these two aspects. We note, however, that our observational results do not rule out the two-step processes identified in Table \ref{Tab:ere_models}.

\section{Summary}
We presented new wide-field digital optical CCD imaging data of IC 63 and IC 59, two nebulae externally illuminated by the early B star $\gamma$ Cas  (B0.5 IVpe). The three photometric bands were chosen such that the first two, ($B$, $G$), covered the scattered light continuum, while the third, $R$, covered parts of the ERE in the 620 -- 690 nm range. Additional images, taken with narrow-band filters centred on the H$\alpha$ and \SII \, emission lines, made possible the removal of contributions from these two emission process to the R-band images. We also secured archival data from \textit{Spitzer}, \textit{Herschel}, and \textit{WISE} space observatories, which permitted the study of the spatial distributions of band emissions from both neutral and ionized PAHs and thermal emissions from large dust grains in IC 59 and IC 63, as well as emissions from pure rotational transitions of UV-excited molecular hydrogen in IC 63. The analysis of these data led to the following conclusions:

1. We detected and mapped the ERE in both nebulae. The band-integrated intensity of the ERE in IC 63 was found to be in excellent agreement with the single previous detection of ERE in IC 63. In IC 59, the maximum measured ERE intensities were consistent with previously reported upper limits.

2. We confirmed illumination geometries with respect to $\gamma$ Cas in which IC 63 is being irradiated at angles near 90$^{\circ}$, producing a PDR seen nearly edge-on. By contrast, IC 59 is located behind the plane of the sky containing $\gamma$ Cas, resulting in scattering angles near 135$^{\circ}$.

3. In IC 63, peak ERE intensities were observed immediately behind the narrow hydrogen ionization front, followed in increasing depth by the peak emissions of first ionized PAHs (\textit{Spitzer} IRAC4) and then PAHs seen by \textit{WISE} W3, which traces both ionized and neutral PAHs, in addition to emission from stochastically heated tiny grains. The closest spatial overlap occurs between ERE and pure rotational emission from H$_{2}$. 

4. Maximum ERE intensities in IC 59 were observed in a layer in front of the PAH emission recorded in the \textit{WISE} W3 band and the \textit{Spitzer} IRAC4 band, consistent with our findings in IC 63. Absolute ERE intensities were almost constant while $R$-band scattered light intensities varied by a factor of five, indicating ERE saturation resulting from an excitation process limited to a narrow energy range below the Lyman limit of atomic hydrogen.

5. An ERE excitation model, applied to the observations of IC 59 displayed in a colour difference diagram, yielded an independent determination of the energy range in which the ERE in IC 59 is excited. The data suggest E$_{\text{exc}}$ >  10.2 eV.

6. The absolute ERE intensities in IC 63 and IC 59 scale well with the FUV radiation field densities due to the illumination by $\gamma$ Cas with a distance ratio of 2.5. However, the photon conversion efficiency in IC 59 and IC 63 per unit FUV photon was found to be lower by about two orders of magnitude than that reported for the diffuse ISM. This suggests that the ERE carriers are marginally stable against photo-fragmentation in the low radiation field densities of the diffuse ISM and are largely destroyed in the $10^{2}$ to $10^{3}$ times higher radiation field densities prevailing in IC 59 and IC 63.

7. We examined previously existing models of potential ERE carriers. Models relying on classical fluorescence/photoluminescence with characteristically small Stokes shifts appear to be ruled out by our observations. Our data provide strong support for models (\citealt{Leger88}; \citealt{Duley09}) that rely on recurrent fluorescence in highly isolated molecules or molecular ions.  These models demand FUV photon excitation and naturally result in the exceptionally high ERE photon conversion efficiencies found in the diffuse ISM. These models have also received strong experimental laboratory support in recent years. We note, however, that our data do not exclude two-step processes, where FUV photons are required to produce the ERE carriers, followed by photo-excitation at longer wavelengths.

\section{Acknowledgments}
The authors thank the anonymous referee for the careful reading and useful feedback. This work used ancillary data from $Spitzer$, $WISE$, and $Herschel$, retrieved from NASA/IPAC Infrared Science Archive. We also thank Don Goldman for providing the original data for Figure 1, and Brian Fleming for providing us with a fits file of H$_2$ rotational line emission, including WCS solution, for IC 63. TSYL gratefully acknowledges support from the \textit{Doreen and Lyman Spitzer Graduate Fellowship} and \textit{the Canaday Family Charitable Trust} for this project. 




\bibliographystyle{mn2e}
\bibliography{ERE} 

\begin{thebibliography}{}

\bibitem[\protect\citeauthoryear{{Alecian}, {Catala}, {Wade}, {Donati},
  {Petit}, {Landstreet}, {B{\"o}hm}, {Bouret}, {Bagnulo}, {Folsom}, {Grunhut}
  \& {Silvester}}{{Alecian} et~al.}{2008}]{Alecian08}
{Alecian} E.,  {Catala} C.,  {Wade} G.~A.,  {Donati} J.-F.,  {Petit} P.,
  {Landstreet} J.~D.,  {B{\"o}hm} T.,  {Bouret} J.-C.,  {Bagnulo} S.,  {Folsom}
  C.,  {Grunhut} J.,    {Silvester} J.,  2008, \mnras, 385, 391

\bibitem[\protect\citeauthoryear{{Allain}, {Leach} \& {Sedlmayr}}{{Allain}
  et~al.}{1996}]{Allain96}
{Allain} T.,  {Leach} S.,    {Sedlmayr} E.,  1996, \aap, 305, 602

\bibitem[\protect\citeauthoryear{{Andersson}, {Piirola}, {De Buizer},
  {Clemens}, {Uomoto}, {Charcos-Llorens}, {Geballe}, {Lazarian}, {Hoang} \&
  {Vornanen}}{{Andersson} et~al.}{2013}]{Andersson13}
{Andersson} B.-G.,  {Piirola} V.,  {De Buizer} J.,  {Clemens} D.~P.,  {Uomoto}
  A.,  {Charcos-Llorens} M.,  {Geballe} T.~R.,  {Lazarian} A.,  {Hoang} T.,
  {Vornanen} T.,  2013, \apj, 775, 84

\bibitem[\protect\citeauthoryear{{Bakes}, {Tielens} \& {Bauschlicher}
  Jr.}{{Bakes} et~al.}{2001}]{Bakes01a}
{Bakes} E.~L.~O.,  {Tielens} A.~G.~G.~M.,    {Bauschlicher} Jr. C.~W.,  2001,
  \apj, 556, 501

\bibitem[\protect\citeauthoryear{{Bakes}, {Tielens}, {Bauschlicher} Jr.,
  {Hudgins} \& {Allamandola}}{{Bakes} et~al.}{2001}]{Bakes01b}
{Bakes} E.~L.~O.,  {Tielens} A.~G.~G.~M.,  {Bauschlicher} Jr. C.~W.,  {Hudgins}
  D.~M.,    {Allamandola} L.~J.,  2001, \apj, 560, 261

\bibitem[\protect\citeauthoryear{{Bern{\'e}}, {Joblin}, {Rapacioli}, {Thomas},
  {Cuillandre} \& {Deville}}{{Bern{\'e}} et~al.}{2008}]{Berne08}
{Bern{\'e}} O.,  {Joblin} C.,  {Rapacioli} M.,  {Thomas} J.,  {Cuillandre}
  J.-C.,    {Deville} Y.,  2008, \aap, 479, L41

\bibitem[\protect\citeauthoryear{{Bessell}}{{Bessell}}{1990}]{Bessell90}
{Bessell} M.~S.,  1990, \pasp, 102, 1181

\bibitem[\protect\citeauthoryear{{Blouin}, {McCutcheon}, {Dewdney}, {Roger},
  {Purton}, {Kester} \& {Bontekoe}}{{Blouin} et~al.}{1997}]{Blouin97}
{Blouin} D.,  {McCutcheon} W.~H.,  {Dewdney} P.~E.,  {Roger} R.~S.,  {Purton}
  C.~R.,  {Kester} D.,    {Bontekoe} T.~R.,  1997, \mnras, 287, 455

\bibitem[\protect\citeauthoryear{{Cardelli}, {Clayton} \& {Mathis}}{{Cardelli}
  et~al.}{1989}]{Cardelli89}
{Cardelli} J.~A.,  {Clayton} G.~C.,    {Mathis} J.~S.,  1989, \apj, 345, 245

\bibitem[\protect\citeauthoryear{{Chandrasekaran}, {Kafle}, {Prabhakaran},
  {Heber}, {Rappaport}, {Rubinstein}, {Schwalm}, {toker} \&
  D.}{{Chandrasekaran} et~al.}{2014}]{Chandrasekaran14}
{Chandrasekaran} V.,  {Kafle} B.,  {Prabhakaran} A.,  {Heber} O.,  {Rappaport}
  M.~L.,  {Rubinstein} H.,  {Schwalm} D.,  {toker} Y.,    D. Z.,  2014, J.
  Phys. Chem. Lett., 5, 4078

\bibitem[\protect\citeauthoryear{{Chang}, {Chen} \& {Kwok}}{{Chang}
  et~al.}{2006}]{Chang06}
{Chang} H.-C.,  {Chen} K.,    {Kwok} S.,  2006, \apjl, 639, L63

\bibitem[\protect\citeauthoryear{{Cohen}, {Anderson}, {Cowley}, {Coyne},
  {Fawley}, {Gull}, {Harlan}, {Herbig}, {Holden}, {Hudson}, {Jakoubek},
  {Johnson}, {Merrill}, {Schiffer}, {Soifer} \& {Zuckerman}}{{Cohen}
  et~al.}{1975}]{Cohen75}
{Cohen} M.,  {Anderson} C.~M.,  {Cowley} A.,  {Coyne} G.~V.,  {Fawley} W.,
  {Gull} T.~R.,  {Harlan} E.~A.,  {Herbig} G.~H.,  {Holden} F.,  {Hudson}
  H.~S.,  {Jakoubek} R.~O.,  {Johnson} H.~M.,  {Merrill} K.~M.,  {Schiffer}
  F.~H.,  {Soifer} B.~T.,    {Zuckerman} B.,  1975, \apj, 196, 179

\bibitem[\protect\citeauthoryear{{Colina}, {Bohlin} \& {Castelli}}{{Colina}
  et~al.}{1996}]{Colina96}
{Colina} L.,  {Bohlin} R.~C.,    {Castelli} F.,  1996, \aj, 112, 307

\bibitem[\protect\citeauthoryear{{Croiset}, {Candian}, {Bern{\'e}} \&
  {Tielens}}{{Croiset} et~al.}{2016}]{Croiset16}
{Croiset} B.~A.,  {Candian} A.,  {Bern{\'e}} O.,    {Tielens} A.~G.~G.~M.,
  2016, \aap, 590, A26

\bibitem[\protect\citeauthoryear{{Darbon}, {Perrin} \& {Sivan}}{{Darbon}
  et~al.}{1999}]{Darbon99}
{Darbon} S.,  {Perrin} J.-M.,    {Sivan} J.-P.,  1999, \aap, 348, 990

\bibitem[\protect\citeauthoryear{{Darbon}, {Zavagno}, {Perrin}, {Savine},
  {Ducci} \& {Sivan}}{{Darbon} et~al.}{2000}]{Darbon00}
{Darbon} S.,  {Zavagno} A.,  {Perrin} J.-M.,  {Savine} C.,  {Ducci} V.,
  {Sivan} J.-P.,  2000, \aap, 364, 723

\bibitem[\protect\citeauthoryear{{D'Hendecourt}, {Leger}, {Olofsson} \&
  {Schmidt}}{{D'Hendecourt} et~al.}{1986}]{D'Hendecourt86}
{D'Hendecourt} L.~B.,  {Leger} A.,  {Olofsson} G.,    {Schmidt} W.,  1986,
  \aap, 170, 91

\bibitem[\protect\citeauthoryear{{Duley}}{{Duley}}{1985}]{Duley85}
{Duley} W.~W.,  1985, \mnras, 215, 259

\bibitem[\protect\citeauthoryear{{Duley}}{{Duley}}{2006}]{Duley06}
{Duley} W.~W.,  2006, Faraday Discussions, 133, 415

\bibitem[\protect\citeauthoryear{{Duley}}{{Duley}}{2009}]{Duley09}
{Duley} W.~W.,  2009, \apj, 705, 446

\bibitem[\protect\citeauthoryear{{Ebara}, {Furukawa}, {Matsumoto}, {Tanuma},
  {Azuma}, {Shiromaru} \& {Hansen}}{{Ebara} et~al.}{2016}]{Ebara16}
{Ebara} Y.,  {Furukawa} T.,  {Matsumoto} J.,  {Tanuma} H.,  {Azuma} T.,
  {Shiromaru} H.,    {Hansen} K.,  2016, Physical Review Letters, 117, 133004

\bibitem[\protect\citeauthoryear{{Fleming}, {France}, {Lupu} \&
  {McCandliss}}{{Fleming} et~al.}{2010}]{Fleming10}
{Fleming} B.,  {France} K.,  {Lupu} R.~E.,    {McCandliss} S.~R.,  2010, \apj,
  725, 159

\bibitem[\protect\citeauthoryear{{France}, {Andersson}, {McCandliss} \&
  {Feldman}}{{France} et~al.}{2005}]{France05}
{France} K.,  {Andersson} B.-G.,  {McCandliss} S.~R.,    {Feldman} P.~D.,
  2005, \apj, 628, 750

\bibitem[\protect\citeauthoryear{{Friedmann}, {Witt}, {Gordon}, {Schild},
  {Bohlin} \& {Stecher}}{{Friedmann} et~al.}{1996}]{Friedmann96}
{Friedmann} B.~C.,  {Witt} A.~N.,  {Gordon} K.~D.,  {Schild} R.~E.,  {Bohlin}
  R.~C.,    {Stecher} T.~P.,  1996, in American Astronomical Society Meeting
  Abstracts Vol.~28 of Bulletin of the American Astronomical Society,
  {Large-Angle Scattering in the UV : IC 63}.
p.~1295

\bibitem[\protect\citeauthoryear{{Furton} \& {Witt}}{{Furton} \&
  {Witt}}{1990}]{Furton90}
{Furton} D.~G.,  {Witt} A.~N.,  1990, \apjl, 364, L45

\bibitem[\protect\citeauthoryear{{Furton} \& {Witt}}{{Furton} \&
  {Witt}}{1992}]{Furton92}
{Furton} D.~G.,  {Witt} A.~N.,  1992, \apj, 386, 587

\bibitem[\protect\citeauthoryear{{Godard} \& {Dartois}}{{Godard} \&
  {Dartois}}{2010}]{Godard10}
{Godard} M.,  {Dartois} E.,  2010, \aap, 519, A39

\bibitem[\protect\citeauthoryear{{Gordon}, {Witt} \& {Friedmann}}{{Gordon}
  et~al.}{1998}]{Gordon98}
{Gordon} K.~D.,  {Witt} A.~N.,    {Friedmann} B.~C.,  1998, \apj, 498, 522

\bibitem[\protect\citeauthoryear{{Habart}, {Boulanger}, {Verstraete},
  {Walmsley} \& {Pineau des For{\^e}ts}}{{Habart} et~al.}{2004}]{Habart04}
{Habart} E.,  {Boulanger} F.,  {Verstraete} L.,  {Walmsley} C.~M.,    {Pineau
  des For{\^e}ts} G.,  2004, \aap, 414, 531

\bibitem[\protect\citeauthoryear{{Hall} \& {Williams}}{{Hall} \&
  {Williams}}{1995}]{Hall95}
{Hall} P.,  {Williams} D.~A.,  1995, \apss, 229, 49

\bibitem[\protect\citeauthoryear{{Henry} \& {Smith}}{{Henry} \&
  {Smith}}{2012}]{Henry12}
{Henry} G.~W.,  {Smith} M.~A.,  2012, \apj, 760, 10

\bibitem[\protect\citeauthoryear{{Hobbs}, {York}, {Snow}, {Oka}, {Thorburn},
  {Bishof}, {Friedman}, {McCall}, {Rachford}, {Sonnentrucker} \&
  {Welty}}{{Hobbs} et~al.}{2008}]{Hobbs08}
{Hobbs} L.~M.,  {York} D.~G.,  {Snow} T.~P.,  {Oka} T.,  {Thorburn} J.~A.,
  {Bishof} M.,  {Friedman} S.~D.,  {McCall} B.~J.,  {Rachford} B.,
  {Sonnentrucker} P.,    {Welty} D.~E.,  2008, \apj, 680, 1256

\bibitem[\protect\citeauthoryear{{Hobbs}, {York}, {Thorburn}, {Snow}, {Bishof},
  {Friedman}, {McCall}, {Oka}, {Rachford}, {Sonnentrucker} \& {Welty}}{{Hobbs}
  et~al.}{2009}]{Hobbs09}
{Hobbs} L.~M.,  {York} D.~G.,  {Thorburn} J.~A.,  {Snow} T.~P.,  {Bishof} M.,
  {Friedman} S.~D.,  {McCall} B.~J.,  {Oka} T.,  {Rachford} B.,
  {Sonnentrucker} P.,    {Welty} D.~E.,  2009, \apj, 705, 32

\bibitem[\protect\citeauthoryear{{Hoyle} \& {Wickramasinghe}}{{Hoyle} \&
  {Wickramasinghe}}{1996}]{Hoyle96}
{Hoyle} F.,  {Wickramasinghe} N.~C.,  1996, \apss, 235, 343

\bibitem[\protect\citeauthoryear{{Hoyle} \& {Wickramasinghe}}{{Hoyle} \&
  {Wickramasinghe}}{1999}]{Hoyle99}
{Hoyle} F.,  {Wickramasinghe} N.~C.,  1999, \apss, 268, 321

\bibitem[\protect\citeauthoryear{{Hurwitz}}{{Hurwitz}}{1998}]{Hurwitz98}
{Hurwitz} M.,  1998, \apjl, 500, L67

\bibitem[\protect\citeauthoryear{{Ito}, {Furukawa}, {Tanuma}, {Matsumoto},
  {Shiromaru}, {Majima}, {Goto}, {Azuma} \& {Hansen}}{{Ito}
  et~al.}{2014}]{Ito14}
{Ito} G.,  {Furukawa} T.,  {Tanuma} H.,  {Matsumoto} J.,  {Shiromaru} H.,
  {Majima} T.,  {Goto} M.,  {Azuma} T.,    {Hansen} K.,  2014, Physical Review
  Letters, 112, 183001

\bibitem[\protect\citeauthoryear{{Jansen}, {van Dishoeck} \& {Black}}{{Jansen}
  et~al.}{1994}]{Jansen94}
{Jansen} D.~J.,  {van Dishoeck} E.~F.,    {Black} J.~H.,  1994, \aap, 282, 605

\bibitem[\protect\citeauthoryear{{Jansen}, {van Dishoeck}, {Black}, {Spaans} \&
  {Sosin}}{{Jansen} et~al.}{1995}]{Jansen95}
{Jansen} D.~J.,  {van Dishoeck} E.~F.,  {Black} J.~H.,  {Spaans} M.,    {Sosin}
  C.,  1995, \aap, 302, 223

\bibitem[\protect\citeauthoryear{{Jansen}, {van Dishoeck}, {Keene}, {Boreiko}
  \& {Betz}}{{Jansen} et~al.}{1996}]{Jansen96}
{Jansen} D.~J.,  {van Dishoeck} E.~F.,  {Keene} J.,  {Boreiko} R.~T.,    {Betz}
  A.~L.,  1996, \aap, 309, 899

\bibitem[\protect\citeauthoryear{{Karr}, {Noriega-Crespo} \& {Martin}}{{Karr}
  et~al.}{2005}]{Karr05}
{Karr} J.~L.,  {Noriega-Crespo} A.,    {Martin} P.~G.,  2005, \aj, 129, 954

\bibitem[\protect\citeauthoryear{{Lakowicz}}{{Lakowicz}}{2006}]{Lakowicz06}
{Lakowicz} J.~R.,  2006, {Principles of Fluorescence Spectroscopy}

\bibitem[\protect\citeauthoryear{{Ledoux}, {Ehbrecht}, {Guillois}, {Huisken},
  {Kohn}, {Laguna}, {Nenner}, {Paillard}, {Papoular}, {Porterat} \&
  {Reynaud}}{{Ledoux} et~al.}{1998}]{Ledoux98}
{Ledoux} G.,  {Ehbrecht} M.,  {Guillois} O.,  {Huisken} F.,  {Kohn} B.,
  {Laguna} M.~A.,  {Nenner} I.,  {Paillard} V.,  {Papoular} R.,  {Porterat} D.,
     {Reynaud} C.,  1998, \aap, 333, L39

\bibitem[\protect\citeauthoryear{{Leger}, {D'Hendecourt} \& {Boissel}}{{Leger}
  et~al.}{1988}]{Leger88}
{Leger} A.,  {D'Hendecourt} L.,    {Boissel} P.,  1988, Physical Review
  Letters, 60, 921

\bibitem[\protect\citeauthoryear{{Luhman}, {Luhman}, {Benedict}, {Jaffe} \&
  {Fischer}}{{Luhman} et~al.}{1997}]{Luhman97}
{Luhman} M.~L.,  {Luhman} K.~L.,  {Benedict} T.,  {Jaffe} D.~T.,    {Fischer}
  J.,  1997, \apjl, 480, L133

\bibitem[\protect\citeauthoryear{{Martin}, {Bernard}, {Br{\'e}dy}, {Concina},
  {Joblin}, {Ji}, {Ortega} \& {Chen}}{{Martin} et~al.}{2013}]{Martin13}
{Martin} S.,  {Bernard} J.,  {Br{\'e}dy} R.,  {Concina} B.,  {Joblin} C.,  {Ji}
  M.,  {Ortega} C.,    {Chen} L.,  2013, Physical Review Letters, 110, 063003

\bibitem[\protect\citeauthoryear{{Matsuoka}, {Ienaka}, {Kawara} \&
  {Oyabu}}{{Matsuoka} et~al.}{2011}]{Matsuoka11}
{Matsuoka} Y.,  {Ienaka} N.,  {Kawara} K.,    {Oyabu} S.,  2011, \apj, 736, 119

\bibitem[\protect\citeauthoryear{{Miao}, {Sugitani}, {White} \&
  {Nelson}}{{Miao} et~al.}{2010}]{Miao10}
{Miao} J.,  {Sugitani} K.,  {White} G.~J.,    {Nelson} R.~P.,  2010, \apj, 717,
  658

\bibitem[\protect\citeauthoryear{{Nitzan} \& {Jortner}}{{Nitzan} \&
  {Jortner}}{1979}]{Nitzan79}
{Nitzan} A.,  {Jortner} J.,  1979, \jcp, 71, 3524

\bibitem[\protect\citeauthoryear{{Perrin} \& {Sivan}}{{Perrin} \&
  {Sivan}}{1992}]{Perrin92}
{Perrin} J.-M.,  {Sivan} J.-P.,  1992, \aap, 255, 271

\bibitem[\protect\citeauthoryear{{Platt}}{{Platt}}{1956}]{Platt56}
{Platt} J.~R.,  1956, \apj, 123, 486

\bibitem[\protect\citeauthoryear{{Pollmann}, {Vollmann} \& {Henry}}{{Pollmann}
  et~al.}{2014}]{Pollmann14}
{Pollmann} E.,  {Vollmann} W.,    {Henry} G.~W.,  2014, Information Bulletin on
  Variable Stars, 6109

\bibitem[\protect\citeauthoryear{{Rhee}, {Lee}, {Gudipati}, {Allamandola} \&
  {Head-Gordon}}{{Rhee} et~al.}{2007}]{Rhee07}
{Rhee} Y.~M.,  {Lee} T.~J.,  {Gudipati} M.~S.,  {Allamandola} L.~J.,
  {Head-Gordon} M.,  2007, Proceedings of the National Academy of Science, 104,
  5274

\bibitem[\protect\citeauthoryear{{Rosi}, {Bauschlicher} Jr. \& {Bakes}}{{Rosi}
  et~al.}{2004}]{Rosi04}
{Rosi} M.,  {Bauschlicher} Jr. C.~W.,    {Bakes} E.~L.~O.,  2004, \apj, 609,
  1192

\bibitem[\protect\citeauthoryear{{Ryter}}{{Ryter}}{1996}]{Ryter96}
{Ryter} C.~E.,  1996, \apss, 236, 285

\bibitem[\protect\citeauthoryear{{Sakata}, {Wada}, {Narisawa}, {Asano},
  {Iijima}, {Onaka} \& {Tokunaga}}{{Sakata} et~al.}{1992}]{Sakata92}
{Sakata} A.,  {Wada} S.,  {Narisawa} T.,  {Asano} Y.,  {Iijima} Y.,  {Onaka}
  T.,    {Tokunaga} A.~T.,  1992, \apjl, 393, L83

\bibitem[\protect\citeauthoryear{{Schmidt}, {Cohen} \& {Margon}}{{Schmidt}
  et~al.}{1980}]{Schmidt80}
{Schmidt} G.~D.,  {Cohen} M.,    {Margon} B.,  1980, \apjl, 239, L133

\bibitem[\protect\citeauthoryear{{Seahra} \& {Duley}}{{Seahra} \&
  {Duley}}{1999}]{Seahra99}
{Seahra} S.~S.,  {Duley} W.~W.,  1999, \apj, 520, 719

\bibitem[\protect\citeauthoryear{{Smith} \& {Witt}}{{Smith} \&
  {Witt}}{2002}]{Smith&Witt02}
{Smith} T.~L.,  {Witt} A.~N.,  2002, \apj, 565, 304

\bibitem[\protect\citeauthoryear{{Stokes}}{{Stokes}}{1852}]{Stokes1852}
{Stokes} G.~G.,  1852, Philosophical Transactions of the Royal Society of
  London Series I, 142, 463

\bibitem[\protect\citeauthoryear{{Szomoru} \& {Guhathakurta}}{{Szomoru} \&
  {Guhathakurta}}{1998}]{Szomoru98}
{Szomoru} A.,  {Guhathakurta} P.,  1998, \apjl, 494, L93

\bibitem[\protect\citeauthoryear{{Telting}, {Waters}, {Persi} \&
  {Dunlop}}{{Telting} et~al.}{1993}]{Telting93}
{Telting} J.~H.,  {Waters} L.~B.~F.~M.,  {Persi} P.,    {Dunlop} S.~R.,  1993,
  \aap, 270, 355

\bibitem[\protect\citeauthoryear{{Thi}, {van Dishoeck}, {Bell}, {Viti} \&
  {Black}}{{Thi} et~al.}{2009}]{Thi09}
{Thi} W.-F.,  {van Dishoeck} E.~F.,  {Bell} T.,  {Viti} S.,    {Black} J.,
  2009, \mnras, 400, 622

\bibitem[\protect\citeauthoryear{{Thompson}, {Parker}, {Day}, {Connor} \&
  {Evans}}{{Thompson} et~al.}{2013}]{Thompson13}
{Thompson} S.~P.,  {Parker} J.~E.,  {Day} S.~J.,  {Connor} L.~D.,    {Evans}
  A.,  2013, \mnras, 434, 2582

\bibitem[\protect\citeauthoryear{{van Leeuwen}}{{van
  Leeuwen}}{2007}]{VanLeeuwen07}
{van Leeuwen} F.,  ed. 2007, {Hipparcos, the New Reduction of the Raw Data}
  Vol.~350 of Astrophysics and Space Science Library

\bibitem[\protect\citeauthoryear{{Verstraete}, {Leger}, {D'Hendecourt},
  {Defourneau} \& {Dutuit}}{{Verstraete} et~al.}{1990}]{Verstraete90}
{Verstraete} L.,  {Leger} A.,  {D'Hendecourt} L.,  {Defourneau} D.,    {Dutuit}
  O.,  1990, \aap, 237, 436

\bibitem[\protect\citeauthoryear{{Vijh}, {Witt} \& {Gordon}}{{Vijh}
  et~al.}{2005}]{Vijh05}
{Vijh} U.~P.,  {Witt} A.~N.,    {Gordon} K.~D.,  2005, \apj, 619, 368

\bibitem[\protect\citeauthoryear{{Vijh}, {Witt}, {York}, {Dwarkadas},
  {Woodgate} \& {Palunas}}{{Vijh} et~al.}{2006}]{Vijh06}
{Vijh} U.~P.,  {Witt} A.~N.,  {York} D.~G.,  {Dwarkadas} V.~V.,  {Woodgate}
  B.~E.,    {Palunas} P.,  2006, \apj, 653, 1336

\bibitem[\protect\citeauthoryear{{Webster}}{{Webster}}{1993}]{Webster93}
{Webster} A.,  1993, \mnras, 264, L1

\bibitem[\protect\citeauthoryear{{Weingartner} \& {Draine}}{{Weingartner} \&
  {Draine}}{2001}]{Weingartner01}
{Weingartner} J.~C.,  {Draine} B.~T.,  2001, \apj, 548, 296

\bibitem[\protect\citeauthoryear{{Witt}}{{Witt}}{1985}]{Witt85}
{Witt} A.~N.,  1985, \apj, 294, 216

\bibitem[\protect\citeauthoryear{{Witt}}{{Witt}}{2014}]{Witt14}
{Witt} A.~N.,  2014, in {Cami} J.,  {Cox} N.~L.~J.,  eds, The Diffuse
  Interstellar Bands Vol.~297 of IAU Symposium, {Blue Luminescence and Extended
  Red Emission: Possible Connections to the Diffuse Interstellar Bands}.
pp 173--179

\bibitem[\protect\citeauthoryear{{Witt} \& {Boroson}}{{Witt} \&
  {Boroson}}{1990}]{Witt90}
{Witt} A.~N.,  {Boroson} T.~A.,  1990, \apj, 355, 182

\bibitem[\protect\citeauthoryear{{Witt}, {Gordon} \& {Furton}}{{Witt}
  et~al.}{1998}]{Witt98}
{Witt} A.~N.,  {Gordon} K.~D.,    {Furton} D.~G.,  1998, \apjl, 501, L111

\bibitem[\protect\citeauthoryear{{Witt}, {Gordon}, {Vijh}, {Sell}, {Smith} \&
  {Xie}}{{Witt} et~al.}{2006}]{Witt06}
{Witt} A.~N.,  {Gordon} K.~D.,  {Vijh} U.~P.,  {Sell} P.~H.,  {Smith} T.~L.,
  {Xie} R.-H.,  2006, \apj, 636, 303

\bibitem[\protect\citeauthoryear{{Witt} \& {Malin}}{{Witt} \&
  {Malin}}{1989}]{Witt89a}
{Witt} A.~N.,  {Malin} D.~F.,  1989, \apjl, 347, L25

\bibitem[\protect\citeauthoryear{{Witt}, {Petersohn}, {Bohlin}, {O'Connell},
  {Roberts}, {Smith} \& {Stecher}}{{Witt} et~al.}{1992}]{Witt92}
{Witt} A.~N.,  {Petersohn} J.~K.,  {Bohlin} R.~C.,  {O'Connell} R.~W.,
  {Roberts} M.~S.,  {Smith} A.~M.,    {Stecher} T.~P.,  1992, \apjl, 395, L5

\bibitem[\protect\citeauthoryear{{Witt} \& {Schild}}{{Witt} \&
  {Schild}}{1988}]{Witt88}
{Witt} A.~N.,  {Schild} R.~E.,  1988, \apj, 325, 837

\bibitem[\protect\citeauthoryear{{Witt}, {Stecher}, {Boroson} \&
  {Bohlin}}{{Witt} et~al.}{1989}]{Witt89}
{Witt} A.~N.,  {Stecher} T.~P.,  {Boroson} T.~A.,    {Bohlin} R.~C.,  1989,
  \apjl, 336, L21

\bibitem[\protect\citeauthoryear{{Witt} \& {Vijh}}{{Witt} \&
  {Vijh}}{2004}]{Witt04}
{Witt} A.~N.,  {Vijh} U.~P.,  2004, in {Witt} A.~N.,  {Clayton} G.~C.,
  {Draine} B.~T.,  eds, Astrophysics of Dust Vol.~309 of Astronomical Society
  of the Pacific Conference Series, {Extended Red Emission: Photoluminescence
  by Interstellar Nanoparticles}.
p.~115

\bibitem[\protect\citeauthoryear{{Witt}, {Walker}, {Bohlin} \&
  {Stecher}}{{Witt} et~al.}{1982}]{Witt82}
{Witt} A.~N.,  {Walker} G.~A.~H.,  {Bohlin} R.~C.,    {Stecher} T.~P.,  1982,
  \apj, 261, 492

\bibitem[\protect\citeauthoryear{{Zhen}, {Rodriguez Castillo}, {Joblin},
  {Mulas}, {Sabbah}, {Giuliani}, {Nahon}, {Martin}, {Champeaux} \&
  {Mayer}}{{Zhen} et~al.}{2016}]{Zhen16}
{Zhen} J.,  {Rodriguez Castillo} S.,  {Joblin} C.,  {Mulas} G.,  {Sabbah} H.,
  {Giuliani} A.,  {Nahon} L.,  {Martin} S.,  {Champeaux} J.-P.,    {Mayer}
  P.~M.,  2016, \apj, 822, 113

\end{thebibliography}



\setcounter{table}{0}
\setcounter{figure}{0}
\renewcommand{\thetable}{A\arabic{table}}
\renewcommand\thefigure{A\arabic{figure}}

\onecolumn
\appendix
\section{Regions with measured surface brightness to flux ratio in IC59 and IC63}

\begin{center}
\begin{longtable}{ccccc}
\caption{Regions with measured surface brightness to flux ratio in IC59} 
\label{Tab:IC59_sbflux} \\

\hline \multicolumn{1}{c}{R.A.} & \multicolumn{1}{c}{Dec.} & \multicolumn{1}{c}{$log \bigg\{\frac{S(B)}{F^{*}(B)} \bigg\}$} & \multicolumn{1}{c}{$log \bigg\{\frac{S(G)}{F^{*}(G)} \bigg\}$} & \multicolumn{1}{c}{$log \bigg\{\frac{S(R_{-})}{F^{*}(R_{-})} \bigg\}$}\\
\multicolumn{1}{c}{(J2000)} & \multicolumn{1}{c}{(J2000)} & \multicolumn{1}{c}{$[sr^{-1}]$} & \multicolumn{1}{c}{$[sr^{-1}]$} & \multicolumn{1}{c}{$[sr^{-1}]$}\\\hline \hline
\endfirsthead

\multicolumn{5}{c}%
{{\tablename\ \thetable{} -- continued from previous page}} \\
\hline \multicolumn{1}{c}{R.A.} & \multicolumn{1}{c}{Dec.} & \multicolumn{1}{c}{$log \bigg\{\frac{S(B)}{F^{*}(B)} \bigg\}$} & \multicolumn{1}{c}{$log \bigg\{\frac{S(G)}{F^{*}(G)} \bigg\}$} & \multicolumn{1}{c}{$log \bigg\{\frac{S(R_{-})}{F^{*}(R_{-})} \bigg\}$}\\
\multicolumn{1}{c}{(J2000)} & \multicolumn{1}{c}{(J2000)} & \multicolumn{1}{c}{$[sr^{-1}]$} & \multicolumn{1}{c}{$[sr^{-1}]$} & \multicolumn{1}{c}{$[sr^{-1}]$}\\\hline \hline
\endhead

\hline \multicolumn{5}{r}{{Continued on next page}} \\ \hline
\endfoot

\hline \hline
\endlastfoot

0:56:20.66 & +61:10:35.03 & 2.08$\pm$0.02 & 2.14$\pm$0.02 & 2.20$\pm$0.02 \\
0:56:28.86 & +61:09:46.26 & 2.12 & 2.16 & 2.18 \\
0:56:37.85 & +61:09:27.34 & 2.09 & 2.13 & 2.13 \\
0:56:41.24 & +61:08:01.79 & 2.16 & 2.20 & 2.22 \\
0:56:42.54 & +61:08:36.10 & 2.13 & 2.18 & 2.22 \\
0:56:45.27 & +61:08:06.96 & 2.12 & 2.16 & 2.21 \\
0:56:48.36 & +61:06:24.52 & 1.88 & 1.86 & 1.97 \\
0:56:49.22 & +61:08:27.47 & 2.00 & 2.02 & 2.07 \\
0:56:49.78 & +61:07:14.95 & 2.11 & 2.12 & 2.16 \\
0:56:53.31 & +61:09:37.04 & 2.17 & 2.21 & 2.25 \\
0:56:53.71 & +61:07:26.32 & 2.21 & 2.23 & 2.27 \\
0:56:54.21 & +61:08:02.33 & 2.05 & 2.09 & 2.16 \\
0:56:54.73 & +61:06:03.67 & 1.99 & 1.97 & 2.06 \\
0:56:59.68 & +61:06:32.49 & 1.92 & 1.93 & 2.04 \\
0:57:00.69 & +61:07:02.37 & 1.95 & 1.94 & 2.02 \\
0:57:01.54 & +61:07:43.24 & 2.29 & 2.33 & 2.37 \\
0:57:02.85 & +61:08:29.55 & 2.16 & 2.19 & 2.20 \\
0:57:04.03 & +61:07:11.21 & 2.01 & 1.97 & 2.09 \\
0:57:05.47 & +61:07:55.63 & 2.27 & 2.32 & 2.35 \\
0:57:05.88 & +61:07:25.50 & 2.28 & 2.31 & 2.38 \\
0:57:06.64 & +61:06:56.98 & 1.92 & 1.88 & 2.05 \\
0:57:10.74 & +61:07:07.04 & 2.23 & 2.20 & 2.32 \\
0:57:12.57 & +61:06:52.54 & 2.01 & 1.95 & 2.06 \\
0:57:14.65 & +61:07:07.17 & 2.34 & 2.34 & 2.44 \\
0:57:14.73 & +61:06:30.86 & 1.90 & 1.82 & 1.98 \\
0:57:15.76 & +61:07:36.31 & 2.37 & 2.42 & 2.49 \\
0:57:16.38 & +61:06:05.22 & 1.86 & 1.82 & 1.94 \\
0:57:16.79 & +61:07:01.00 & 2.30 & 2.31 & 2.38 \\
0:57:17.05 & +61:08:52.45 & 2.10 & 2.13 & 2.20 \\
0:57:21.10 & +61:05:51.90 & 1.90 & 1.82 & 1.94 \\
0:57:21.85 & +61:07:14.87 & 2.37 & 2.42 & 2.50 \\
0:57:21.98 & +61:06:09.92 & 2.04 & 1.98 & 2.13 \\
0:57:23.50 & +61:06:25.74 & 2.22 & 2.22 & 2.30 \\
0:57:24.40 & +61:05:39.57 & 2.00 & 1.97 & 2.19 \\
0:57:26.95 & +61:05:08.42 & 1.99 & 1.97 & 2.13 \\
0:57:27.69 & +61:05:35.52 & 2.22 & 2.21 & 2.26 \\
0:57:28.17 & +61:04:48.07 & 1.91 & 1.95 & 2.12 \\
0:57:29.07 & +61:06:08.27 & 2.24 & 2.24 & 2.34 \\
0:57:30.01 & +61:05:46.54 & 2.26 & 2.24 & 2.31 \\
0:57:31.89 & +61:07:37.13 & 2.24 & 2.28 & 2.37 \\
0:57:32.53 & +61:05:21.22 & 2.27 & 2.28 & 2.39 \\
0:57:32.54 & +61:09:09.68 & 2.26 & 2.31 & 2.41 \\
0:57:33.25 & +61:05:01.55 & 2.33 & 2.34 & 2.45 \\
0:57:34.46 & +61:07:13.46 & 2.30 & 2.34 & 2.44 \\
0:57:35.44 & +61:08:01.58 & 2.25 & 2.30 & 2.40 \\
0:57:35.50 & +61:06:52.75 & 2.34 & 2.39 & 2.48 \\
0:57:35.67 & +61:04:44.82 & 1.93 & 1.90 & 2.09 \\
0:57:35.70 & +61:05:27.08 & 2.27 & 2.30 & 2.42 \\
0:57:37.05 & +61:10:23.43 & 2.12 & 2.16 & 2.27 \\
0:57:37.85 & +61:08:06.22 & 2.33 & 2.39 & 2.50 \\
0:57:37.93 & +61:06:17.83 & 2.40 & 2.45 & 2.55 \\
0:57:38.68 & +61:05:36.11 & 2.20 & 2.20 & 2.36 \\
0:57:38.93 & +61:08:33.52 & 2.25 & 2.32 & 2.44 \\
0:57:39.31 & +61:04:34.11 & 1.84 & 1.89 & 2.01 \\
0:57:39.46 & +61:05:13.62 & 2.26 & 2.29 & 2.38 \\
0:57:40.38 & +61:10:02.22 & 2.15 & 2.20 & 2.34 \\
0:57:40.69 & +61:05:50.23 & 2.21 & 2.22 & 2.33 \\
0:57:40.95 & +61:06:31.96 & 2.25 & 2.27 & 2.38 \\
0:57:41.12 & +61:07:02.67 & 2.27 & 2.31 & 2.39 \\
0:57:41.38 & +61:08:23.22 & 2.26 & 2.34 & 2.45 \\
0:57:41.70 & +61:07:16.93 & 2.27 & 2.31 & 2.40 \\
0:57:41.77 & +61:04:56.48 & 1.95 & 1.94 & 2.09 \\
0:57:42.10 & +61:05:36.93 & 2.00 & 1.97 & 2.15 \\
0:57:42.39 & +61:07:32.44 & 2.23 & 2.27 & 2.34 \\
0:57:42.67 & +61:06:19.50 & 2.12 & 2.11 & 2.22 \\
0:57:42.78 & +61:06:52.82 & 2.21 & 2.24 & 2.34 \\
0:57:42.87 & +61:05:52.57 & 2.07 & 2.06 & 2.20 \\
0:57:43.03 & +61:09:51.18 & 2.11 & 2.14 & 2.28 \\
0:57:43.97 & +61:05:01.77 & 1.84 & 1.84 & 2.02 \\
0:57:44.02 & +61:07:06.10 & 2.22 & 2.25 & 2.32 \\
0:57:44.36 & +61:08:05.22 & 2.20 & 2.24 & 2.31 \\
0:57:44.67 & +61:07:39.68 & 2.10 & 2.13 & 2.21 \\
0:57:44.68 & +61:05:38.52 & 1.85 & 1.86 & 2.09 \\
0:57:45.39 & +61:06:04.64 & 1.91 & 1.87 & 2.02 \\
0:57:45.49 & +61:08:41.87 & 2.25 & 2.31 & 2.43 \\
0:57:46.03 & +61:06:44.38 & 1.91 & 1.89 & 2.06 \\
0:57:46.80 & +61:08:07.07 & 2.23 & 2.25 & 2.34 \\
0:57:47.03 & +61:10:02.49 & 2.04 & 2.06 & 2.19 \\
0:57:47.05 & +61:06:27.56 & 1.88 & 1.85 & 2.03 \\
0:57:47.79 & +61:07:50.32 & 2.03 & 2.03 & 2.15 \\
0:57:48.37 & +61:08:38.76 & 2.21 & 2.26 & 2.38 \\
0:57:49.35 & +61:07:24.86 & 1.82 & 1.76 & 1.93 \\
0:57:49.38 & +61:08:10.72 & 2.20 & 2.24 & 2.35 \\
0:57:49.66 & +61:08:57.50 & 2.15 & 2.16 & 2.26 \\
0:57:50.80 & +61:07:48.51 & 1.84 & 1.84 & 2.00 \\
0:57:52.17 & +61:08:12.78 & 1.93 & 1.90 & 2.05 \\
0:57:52.76 & +61:08:30.82 & 1.94 & 1.93 & 2.04 \\
0:57:53.01 & +61:08:51.34 & 2.01 & 1.98 & 2.07 \\
\end{longtable}
\end{center}

\newpage
\begin{center}
\begin{longtable}{ccccc}
\caption{Regions with measured surface brightness to flux ratio in IC63} 
\label{Tab:IC63_sbflux} \\

\hline \multicolumn{1}{c}{R.A.} & \multicolumn{1}{c}{Dec.} & \multicolumn{1}{c}{$log \bigg\{\frac{S(B)}{F^{*}(B)} \bigg\}$} & \multicolumn{1}{c}{$log \bigg\{\frac{S(G)}{F^{*}(G)} \bigg\}$} & \multicolumn{1}{c}{$log \bigg\{\frac{S(R_{-})}{F^{*}(R_{-})} \bigg\}$}\\
\multicolumn{1}{c}{(J2000)} & \multicolumn{1}{c}{(J2000)} & \multicolumn{1}{c}{$[sr^{-1}]$} & \multicolumn{1}{c}{$[sr^{-1}]$} & \multicolumn{1}{c}{$[sr^{-1}]$}\\\hline \hline
\endfirsthead

\multicolumn{5}{c}
{{\tablename\ \thetable{} -- continued from previous page}} \\
\hline \multicolumn{1}{c}{R.A.} & \multicolumn{1}{c}{Dec.} & \multicolumn{1}{c}{$log \bigg\{\frac{S(B)}{F^{*}(B)} \bigg\}$} & \multicolumn{1}{c}{$log \bigg\{\frac{S(G)}{F^{*}(G)} \bigg\}$} & \multicolumn{1}{c}{$log \bigg\{\frac{S(R_{-})}{F^{*}(R_{-})} \bigg\}$}\\
\multicolumn{1}{c}{(J2000)} & \multicolumn{1}{c}{(J2000)} & \multicolumn{1}{c}{$[sr^{-1}]$} & \multicolumn{1}{c}{$[sr^{-1}]$} & \multicolumn{1}{c}{$[sr^{-1}]$}\\\hline \hline
\endhead

\hline \multicolumn{5}{r}{{Continued on next page}} \\ \hline
\endfoot

\hline \hline
\endlastfoot
0:58:59.18 & +60:53:15.93 & 2.83$\pm$0.01 & 2.90$\pm$0.01 & 3.07$\pm$0.02 \\
0:59:00.13 & +60:53:46.97 & 2.74 & 2.85 & 3.09 \\
0:59:00.67 & +60:53:24.96 & 2.82 & 2.96 & 3.16 \\
0:59:02.56 & +60:53:06.05 & 2.87 & 2.99 & 3.21 \\
0:59:05.92 & +60:53:48.12 & 2.62 & 2.72 & 2.83 \\
0:59:06.69 & +60:54:27.49 & 2.29 & 2.31 & 2.31 \\
0:59:08.97 & +60:53:15.20 & 2.65 & 2.71 & 2.87 \\
0:59:09.31 & +60:54:19.12 & 2.45 & 2.56 & 2.68 \\
0:59:12.51 & +60:54:21.55 & 2.37 & 2.46 & 2.55 \\
0:59:13.23 & +60:55:05.60 & 2.50 & 2.58 & 2.72 \\
0:59:14.13 & +60:53:29.98 & 2.68 & 2.79 & 3.03 \\
0:59:14.37 & +60:53:58.49 & 2.38 & 2.43 & 2.46 \\
0:59:15.88 & +60:53:20.12 & 2.66 & 2.75 & 2.93 \\
0:59:17.85 & +60:53:14.93 & 2.55 & 2.61 & 2.71 \\
0:59:18.31 & +60:54:26.78 & 2.17 & 2.20 & 2.22 \\
0:59:19.91 & +60:53:58.74 & 2.52 & 2.65 & 2.81 \\
0:59:23.24 & +60:53:21.23 & 2.18 & 2.18 & 2.18 \\
0:59:24.11 & +60:54:33.30 & 2.31 & 2.42 & 2.55 \\
0:59:27.53 & +60:58:14.12 & 2.24 & 2.21 & 2.19 \\
0:59:27.73 & +60:59:10.06 & 2.22 & 2.21 & 2.18 \\
0:59:29.86 & +60:53:03.27 & 2.46 & 2.46 & 2.50 \\
0:59:29.88 & +60:56:05.99 & 2.16 & 2.26 & 2.35 \\
0:59:33.31 & +60:56:34.89 & 2.16 & 2.26 & 2.35 \\
0:59:35.82 & +60:52:46.21 & 2.47 & 2.50 & 2.52 \\
0:59:36.48 & +60:54:47.50 & 2.03 & 2.11 & 2.20 \\
0:59:36.64 & +60:52:30.59 & 2.51 & 2.54 & 2.60 \\
0:59:44.62 & +60:55:33.16 & 2.04 & 2.12 & 2.22 \\
0:59:54.15 & +60:53:57.15 & 2.11 & 2.13 & 2.11 \\
0:59:57.57 & +60:51:58.46 & 2.14 & 2.15 & 2.14 \\
\end{longtable}
\end{center}

\section{IC63 selected regions}
\newpage
\begin{figure}
  \centering
  \includegraphics[width=0.8\columnwidth]{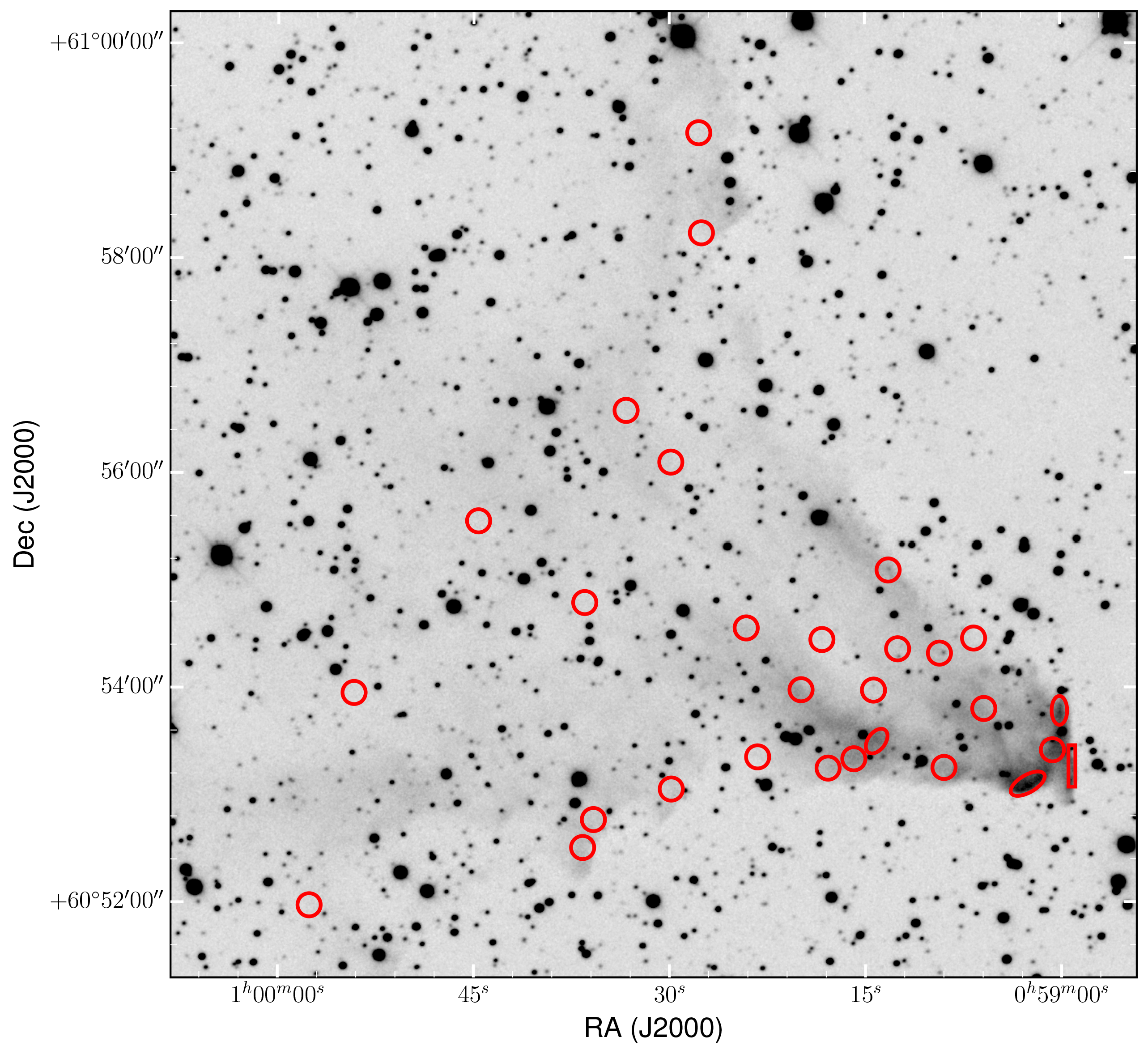}
  \caption{Star-free nebular regions in IC63. Each circle has a radius 6.5\arcsec, except for the five brightest ERE regions, which were measured with apertures according to the shapes of the ERE filaments.}
  \label{Fig:Fig_IC63region}
\end{figure}

\bsp	
\label{lastpage}
\end{document}